\documentclass[useAMS,usenatbib]{mnras}
\usepackage{graphicx,psfig,cite,epsf}  
\usepackage{times}
\usepackage{subfigure}
\usepackage{multirow}
\usepackage{amsmath}	
\usepackage{amssymb}	
\usepackage{gensymb}
\usepackage{colortbl}
\usepackage[utf8]{inputenc}
\usepackage{newtxtext,newtxmath}
\usepackage{color}
\usepackage{graphicx}
\usepackage{natbib}
\usepackage{cite}
\usepackage{geometry}
\usepackage{ae,aecompl}
\usepackage{longtable}
\usepackage{soul}
\usepackage{hyperref} 
\usepackage{lscape}
\def\xmm{{\it XMM-Newton}}
\def\chandra{{\it Chandra}}

\title[ULXs in the Cartwheel]{The largest bright ULX population in a galaxy: X-ray variability and Luminosity Function in the Cartwheel ring Galaxy}

\author[C. Salvaggio et al.]{
Chiara Salvaggio,$^{1,2}$\thanks{E-mail: chiara.salvaggio@inaf.it}
A. Wolter,$^{2}$
A. Belfiore,$^{3}$
M. Colpi$^{1}$
\\
$^{1}$Dipartimento di Fisica, Universit\`a degli Studi di Milano - Bicocca, Piazza della Scienza 3, 20126 Milano, Italy\\
$^{2}$INAF - Osservatorio Astronomico di Brera, via Brera 28, 20121 Milano, Italy\\
$^{3}$INAF - Istituto di Astrofisica Spaziale e Fisica Cosmica di Milano, via A. Corti 12, 20133 Milano, Italy
}

\date{Accepted XXX. Received YYY; in original form ZZZ}

\pubyear{2023}

\begin{document}
\label{firstpage}
\pagerange{\pageref{firstpage}--\pageref{lastpage}}
\maketitle

\begin{abstract}
We analyse all the available \chandra\ observations of the Cartwheel Galaxy and its compact group, taken between 2001 and 2008, with the main aim of addressing the variability in the X-ray band for this spectacular collisional ring galaxy. We focus on the study of point-like sources, in particular we are interested in Ultraluminous X-ray sources (ULXs, $L_{\rm x} \geq$10$^{39}$ erg s$^{-1}$), that we treat as a class.  We exploit archival \xmm\ data to enrich the study of the long-term variability, on timescales of months to years.
We find a total of 44 sources in the group area, of which 37 in total are ULXs positionally linked with the galaxies and of which we can study variability. They are 29 in the Cartwheel itself, 7 in G1 and 1 in G3. About one third of these 37  
sources show long-term variability, while no variability is detected within the single observations. 
Of those, 5 ULXs have a transient behaviour with a maximum range of variability ($L_{\rm max}$/$L_{\rm min}$) of about one order of magnitude and are the best candidate neutron stars.
The X-ray Luminosity Function (XLF) of the point-like sources remains consistent in shape between the \chandra\ observations both for the Cartwheel galaxy itself and for G1, suggesting that flux variability does not strongly influence the average properties of the population on the observation timescales. 

\end{abstract}




\section{Introduction}
\label{sec:intro}
Ultraluminous X-ray sources (ULXs) are point-like, extragalactic and off-nuclear sources, with persistent luminosity in excess of 10$^{39}$ erg s$^{-1}$, which is the Eddington luminosity for accretion onto a $\sim$10 M$_{\odot}$ black hole (BH) (e.g. \citealt{Fabbiano2006}, see also \citealt{Kaaret2017} for a review). 
The off-nuclear constrain excludes the supermassive BH at the center of the galaxy from the census of ULXs.

There is now a consensus on the interpretation of ULXs as accreting compact objects in binary systems (e.g. \citealt{motch2014}), even if the nature of the compact object has been confirmed just for a few sources. The system may be either accreting at super-Eddington rates, in the case of a neutron star (NS) or a stellar mass BH, or at sub-Eddington rate for an intermediate mass BH (masses in the 100 -- 10$^{6}$ $M_{\odot}$ range).

Flux variability is observed in ULXs on different time-scales. Both short-term variability (minutes) (e.g. \citealt{Roberts2003}) and variability on long time-scales (months to years) (e.g. \citealt{Makishima2000,Fabbiano2003}; for the Cartwheel galaxy: \citealt{Wolter2006,Crivellari2009}) have been often taken as an evidence of accretion onto a compact object.  
ULXs may be persistent, albeit with flux variations, or transient.
We consider here as a transient a source detected at least once above 10$^{39}$ erg s$^{-1}$ and once below that threshold, be it as a detection or a significant upper limit. Examples of transient ULXs are M51 ULX-7 \citep{Terashima2004}, CXOU 133705.1-295207 in M83 \citep{Soria2012}, NGC 5907 ULX-2 \citep{Pintore2018} or NGC 55 ULX-2 \citep{Robba2022}. Their outburst may derive from an unstable mass transfer from the companion star (e.g. \citealt{Soria2012,Pintore2018}).

 Coherent pulsations, that can be produced only by a highly magnetised NS, have been measured in at least six ULXs (see \citealt{Bachetti2014,Israel2017,israel2017b,Fuerst2016, Carpano2018,Sathyaprakash2019,Rodriguez2020}), or possibly seven \citep{Quintin2021}. They are indicated as PULXs (pulsar ULXs). PULXs often have large flux variations in their light-curves and/or a transient behaviour and sometimes have a bi-modal flux distribution (e.g. \citealt{Walton2015,motch2014}). The latter may be produced by a propeller phase, i.e. an inhibition of the accretion caused by the magnetosphere of a NS (e.g. \citealt{Illarionov1975,Tsygankov2016a}). 
 Also the detection of a cyclotron line indicates the presence of a NS \citep[e.g.][]{Brightman2018}. 
 
 On the other hand, no methods are currently available to confirm the presence of a BH in a ULX. Indeed, the derivation of the compact object mass through dynamical studies, based on the velocity curves of the optical emission from the companion star, is usually not applicable to ULXs, because optical emission is dominated by X-ray reprocessing in the accretion disc (e.g. \citealt{Roberts2011,Fabrika2021}) and when attempted has given unconfirmed results. Some authors (e.g. \citealt{Koliopanos2017,Pintore2017,Walton2018a}) proposed that NSs may constitute a large fraction of the observed ULX population. \citet{Wiktorowicz2019} compared the number of NSs and BHs expected in the observed and intrinsic ULX population, through simulations of stellar evolution. The NSs are expected to dominate the intrinsic ULX population in regions with solar metallicity, a constant star formation rate (SFR) and an age $>$ 1 Gyr. In contrast, BHs are expected to dominate both the intrinsic and the observed ULX populations at lower metallicity. In case of burst Star Formation (SF), young ULX are dominated by BHs, but for older populations, i.e. post SF, NSs dominate both the intrinsic and observed number of ULXs. 
\newline
\newline
Both in case of a NS or a stellar mass BH, the accretion in ULXs is thought to be mediated by an accretion disc, which is expected to be a supercritical disc with advection and outflows/winds (\citealt{Shakura1973,Poutanen2007}). In this context, the two spectral components usually observed in high statistics ULXs may model the emission from the inner and hot regions of the accretion disc (hard component) and the softer emission from the photosphere of the disc wind (e.g. \citealt{MIddleton2011a}). The short-term aperiodic variability (time-scales from seconds to hours) sometimes observed in ULXs 
(e.g. \citealt{Heil2009,Sutton2013}) can derive from the non-uniform disc winds surrounding ULXs (e.g. \citealt{middleton2015}). The direct evidence of the existence of these winds is shown by the detection of blue-shifted absorption lines in the high resolution, high statistics grating spectra of a few sources,  
(e.g. \citealt{pinto2016,Kosec2018a,Kosec2021}), while indirect evidences are the nebulae observed around a few ULXs in the optical (see e.g. \citealt{Pakull2010}), radio (e.g. \citealt{cseh2012}) or X-ray \citep{belfiore2019} bands.
The aperiodic variability is also observed on long time-scales, i.e. weeks-years, (e.g. M83 ULX, \citealt{Soria2012}; Circinus ULX-5, \citealt{Walton2013}, \citealt{Mondal2021}; ULX-1 and ULX-2 in NGC 925, \citealt{Salvaggio2022}) and may be linked to changes in the mass accretion rate, which also influence the amount and the geometry of the winds ejected from the accretion disc, possibly producing spectral variability.

In addition to non-periodic variations,  
a number of ULXs show periodic flux variability (e.g. NGC 5907 ULX-1, \citealt{Walton2016}; M51 ULX7, \citealt{Vasilopoulos2020,Vasilopoulos2021,Brightman2020}; NGC 925 ULX-3, \citealt{Salvaggio2022,Earnshaw2022}). A periodicity longer than the orbital period (time-scales of tens of days or months) might be present: this super-orbital periodicity is often found in PULXs (e.g. \citealt{Brightman2019}), but we can not rule out that also BH ULXs have this kind of periodicity. By now, how super-orbital periods originate is not clear. One possible explanation is a precession of the accretion disc which obscures the central source (e.g. \citealt{motch2014}); another possibility is a Lense-Thirring precession of the accretion flow (e.g. \citealt{Middleton2018}). 
\newline
\newline
Here,  we address the study of a population of ULXs in the Cartwheel Galaxy  
-- the archetypal collisional ring galaxy -- and in the companion galaxies  G1, G2, G3 that together form a compact group of 4 members \citep[SGC 0035-3357;][]{Iovino2002}. The redshift is $z$ = 0.03 \citep{Amram1998} and the corresponding luminosity distance is 131 Mpc\footnote{Values between $\sim$ 129 and $\sim$ 133 Mpc are indicated in NED for the distance of the Cartwheel galaxy (https://ned.ipac.caltech.edu); we adopt here an intermediate
value. Previous works have used a slightly smaller value of 122 Mpc \citep{Wolter2004}. While comparing our results with previous works, we take into account the different distance assumed. }. At the adopted distance 1 arcsec corresponds to $\sim$ 600 pc. 

The Cartwheel Galaxy X-ray emission has been observed with \textit{Einstein} in 1979, \textit{Rosat} in 1994, \chandra\ in 2001 and 2008, \xmm\ in 2004 and 2005 (\citealt{Wolter1999,Wolter2004,Crivellari2009}).  The very high X-ray activity of the Cartwheel Galaxy is connected with the impact of $\sim 300$ Myr ago, that created the ring galaxy and triggered the episode of star formation (e.g.\citealt{Wolter1999, Higdon1996,mapelli2008}). 
The galaxy contains the largest number of ULXs for a single galaxy (\citealt{Wolter2004, gao2003}), with the only possible exception of NGC 2207. \citet{Mineo2014} listed 28 ULXs from both NGC 2207 and the interacting galaxy IC 2163, while \citet{Walton2022} indicate the presence of 34 ULXs, which however might reduce to just 7 ULXs depending on the real distance of the galaxy complex\footnote{Distances between $\sim$ 13 and $\sim$ 40 Mpc have been estimated for NGC 2207 -- IC 2163 (https://ned.ipac.caltech.edu/).}. However, the sources in NGC 2207 are about a factor of 10 fainter than the Cartwheel ones, even considering the largest estimate for the distance of NGC 2207. 
Most of the Cartwheel ULXs are found in the southern part of the outer ring, co-spatial with the locus of higher star formation, in agreement with the hydrodynamical simulations of \citet{Renaud2018}. Simulations in fact predict more star formation in the portion of the ring furthest from the nucleus, as is observed: the S-SW portion of the outer ring contains also massive and luminous HII regions \citep{Higdon1995}.

The large number of bright ULX sources and the fact that they have been likely created in a single burst of star formation (as assumed also in \citealt{wolter2018}), rendering them coeval, make the Cartwheel the ideal place to apply a population study of ULXs.

The main features of the high energy emission have already been discussed: the first X-ray detection of the Cartwheel ring is presented in \citet{Wolter1999}, the study of the point sources and the X-ray Luminosity Function (XLF) is reported in the work of \citet{Wolter2004} (using the first \chandra\ observation of 2001, published also by \citealt{gao2003}); the extension of that work, using \xmm\ data, is presented in \citet{Crivellari2009}, which addresses also the companion galaxies in detail; the variability of the most luminous source, N10, is studied in \citet{Wolter2006} and \citet{Pizzolato2010}.

This paper completes the analysis of archival data, by addressing the variability of the ULXs as a class, exploiting all the \chandra\ and \xmm\ pointings. We consider also the X-ray properties of the companion galaxies. Given that just three epochs, or five when a source is resolved also by \textit{XMM-Newton}, with a time separation of months/years are available, the data are not suitable for the search of periodicities, instead we focus on the study of long-term aperiodic variability, linked to the accretion rate and flow. The long-term variability is also useful to estimate the fraction of  candidate PULXs, to help characterise the composition of the population (e.g. \citealt{Earnshaw2018,Song2020}). In addition, we analyse the individual \chandra\ observations to investigate the short-term behaviour. We also attempt a spectral analysis when possible (the statistics is usually scanty). We construct the X-ray luminosity function based on \chandra\ data, comparing its shape with that found for other ULX or X-ray binary samples in previous works, and study it at the three different epochs available.

The paper is organized as follows: in Section \ref{sec:datared} we report the observations analysed and the data reduction; in Section \ref{sec:analysis} we describe the data analysis. In Section \ref{sec:results} we present the results of our analysis, while in Sections \ref{sec:discussion} and \ref{sec:conclusion} we discuss the results and give our conclusions.

\section{The Observations}
\label{sec:datared}
\subsection{Chandra Data}
The Cartwheel galaxy has been observed with \chandra\ ACIS-S three times: once in 2001 (obsID 2019, PI: A. Wolter) and twice in 2008 (obsIDs: 9531, 9807, PI: A. Wolter). In exposure 9531 there is a very short period with a background flare, which we exclude from the analysis for cleanliness.  The first \chandra\ exposure is more sensitive than the others in the soft band, owing to a decrease in sensitivity over the years due to the \chandra\ contamination layer. The sensitivity in the total band is less affected. We use the {\sc ciao} task {\sc lim\_sens} to determine the limiting flux in the \chandra\ epochs, which corresponds to a (0.5-10) keV luminosity of $\sim$ 1.5$\times$10$^{39}$ erg s$^{-1}$ in observation 2019 and $\sim$ 2.9$\times$10$^{39}$ erg s$^{-1}$ in the other two observations, assuming a distance of 131 Mpc.  
We consider these differences in response and sensitivity among the epochs in the analysis. 
This sensitivity limit implies that the detected sources are in the ultraluminous regime. 
The journal of the observations used in this work is reported in Table~\ref{chandra_xmm_obs}.

We use standard procedures with the version 4.11 of {\sc ciao}\footnote{http://cxc.harvard.edu/ciao/ \citep{Fruscione2006}} and the calibrations contained in the version 4.8.3 of {\sc caldb}\footnote{https://cxc.cfa.harvard.edu/caldb/downloads/Release\_notes/CALDB\_v4.8.3.html}.  
We re-process all the data with the same procedure for homogeneous results, by using the {\sc ciao} task {\sc chandra\_repro}. 

We perform the spectral analysis with version 12.10.1 of {\sc xspec}\footnote{https://heasarc.gsfc.nasa.gov/xanadu/xspec/ } \citep{Arnaud1996}. We extract the spectra with the {\sc ciao} task {\sc specextract}, in circular regions of 1 arcsecond radius. The background is extracted from an
elliptical annulus around each source (see also appendix \ref{appendixA}).\\

\begin{table}
\begin{center} 
\begin{tabular}{cccc} \hline
Instrument &  obsId &  Date &  T (ks)\\ 
\hline
{\em Chandra} & 2019 & 26/27-May-2001 & 76.1\\
{\em Chandra} & 9531 & 21-Jan-2008 & 51.4 (49.9)\\
 {\em Chandra}& 9807 & 09-Sep-2008 & 49.5\\
{\em XMM-Newton} & 0200800101 & 14-Dec-2004 & 37.2 (25.4)\\
{\em XMM-Newton} & 0200800201 & 21-May-2005 & 60.9 (42.8)\\
\hline
\end{tabular}
\end{center}
\caption{Journal of observations of the Cartwheel Galaxy with the
\chandra\ and \xmm\ satellites. For \xmm\ and \chandra\ observation 9531, we report in parenthesis the net
exposure time left after the removal of high background periods.}
\label{chandra_xmm_obs}
\end{table}
\subsection{XMM-Newton Data}
\label{sec:xmmdata}
The Cartwheel galaxy has been observed with the EPIC instrument on-board \xmm\ in 2004 (obsId: 0200800101, PI: A. Wolter) and in 2005 (obsId: 0200800201, PI: A. Wolter).

The \xmm\ data have also been analysed in \citet{Crivellari2009}. We re-analyse here the data with a more recent software version. 
We reduce data with the Scientific Analysis Software \footnote{https://www.cosmos.esa.int/web/xmm-newton/how-to-use-sas} ({\sc sas}) v.19.0, selecting events with FLAG=0, and PATTERN$\leq$4, for the pn, and FLAG=0, PATTERN$\leq$12, for the MOS detectors. We extract the light curves of high energy photons (energy $>$ 10 keV) to identify periods of high particle background and we exclude the time intervals in which the count rate is $>$ 1 cts s$^{-1}$.

With the task {\sc evselect}\footnote{http://xmm-tools.cosmos.esa.int/external/sas/current/doc/evselect/}, we combine the events of the first and second exposure to create a merged image for each of the EPIC instruments, i.e. pn, MOS1 and MOS2, which we use to do a source detection. The source detection is run with {\sc edetect\_chain}\footnote{https://xmm-tools.cosmos.esa.int/external/sas/current/doc/edetect\_chain/}. We find all the sources already detected in \citet{Crivellari2009}. Thus, to extract the source counts for the \xmm\ sources, we take the coordinates from \citet{Crivellari2009}, who also list the corresponding \chandra\ counterpart for each \xmm\ source. 

When possible, we include the \xmm\ data in the study of the point-sources variability. We just use the higher statistics pn data, because the differences in the calibration among the pn and MOS cameras would give larger uncertainties in the results than those coming from the use of pn data only.

\section{Data analysis}
\label{sec:analysis}
\subsection{Point Sources}
\label{sec:pointsrc}

We detect sources in each \chandra\ observation using the task {\sc wavdetect} (described in the {\sc ciao} threads, using a detection threshold of 2$\times$10$^{-6}$,  
 calculated as one over the number of pixels in the image).  
For source detection we use the energy band (0.3-10) keV, while fluxes and luminosities are computed in the (0.5-10) keV band. The luminosity values indicated in the paper are all unabsorbed values. 
Both the images and the point spread function (PSF) map have been created with the {\sc fluximage} script \footnote{https://cxc.cfa.harvard.edu/ciao/ahelp/fluximage.html}. 
 The detection is made on scales of 1, 2, 4, 8, 16 pixels. Count-rates are small enough not to yield any pile-up. 

 We check for extension using the {\sc ciao} task {\sc srcextent}: all the detected sources are consistent with a point-source at the resolution of the instrument, which, at the Cartwheel distance, corresponds to a radius of $\sim$ 600 pc, i.e. a non-trivial volume of the galaxy.

In order to detect fainter sources not visible in the single observation, we repeat the same procedure on the merged image. To construct it, we reproject the single images with the task {\sc reproject\_obs} and merge them using the {\sc ciao} tool {\sc merge\_obs}.
For the merged image we create a PSF map for each observation and sum them with {\sc dmimgcalc}, weighting for the exposure time. 

We comment the results of the detection in section \ref{sec:poinsrc-res}, while details on the nomenclature, with the list of sources, the consistency checks for coordinates in different datasets and the small incidence of spurious sources are in appendix \ref{appendixA}.

\begin{figure*}
    \centering
    
    \includegraphics[scale=0.60]{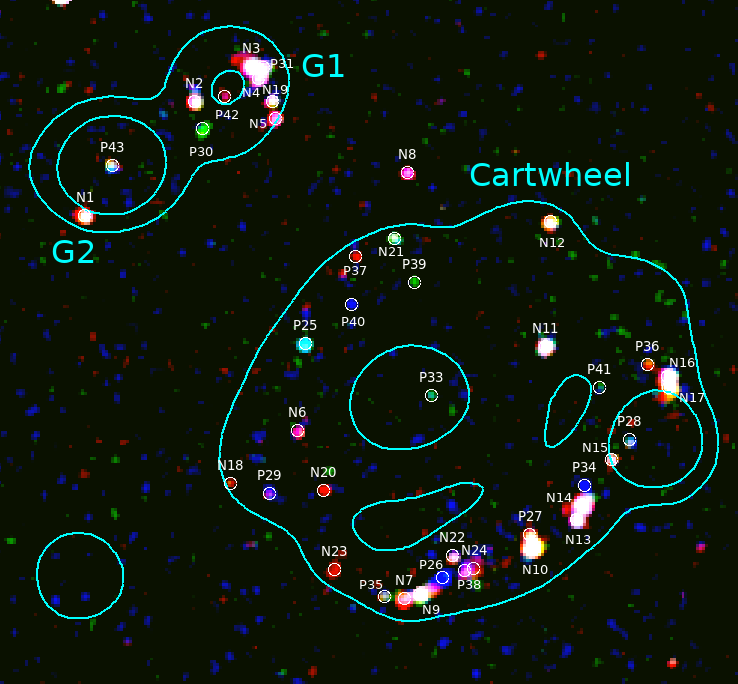}

    \caption{A "time" image of the Cartwheel Galaxy (the large galaxy in the center of the figure) and the companions G1 (on the top to the right) and G2 (on the top to the left). The image is in the energy band 0.3-10 keV and it is exposure corrected. It is a combination of the smoothed \chandra\ data from the three epochs to highlight variability $-$ each data-set in a different color (May 2001 = red, Jan 2008 = green, Sep 2008 = blue). The names of the detected sources are indicated. We superpose smoothed optical contours from the DSS for reference. The field of view is $\sim$ 2$\times$1.9 arcmin. North is up and East to the left. } \label{fig:img_cart}
\end{figure*}

In order to highlight variability in the \chandra\ observations, we construct an image combining the three \chandra\ epochs with {\sc ds9} \citep{Joye2011}. The three images, in the energy band 0.3-10 keV, are exposure corrected. We assign a color to each epoch, 2019 in red, 9531 in green and 9807 in blue, and construct Figure \ref{fig:img_cart} by smoothing the data with the {\sc ciao} tool {\sc csmooth} (using a Gaussian with $\sigma_{min}$=3, $\sigma_{max}$=5). In the field of view considered, the Cartwheel, G1 and G2 are visible. It is readily evident from the different colours of the sources that variability plays a big role in a few years time-frame.

In order to consider also the \xmm\ data, we need to stress that a number of ULXs are not spatially 
resolved, i.e. N7 and N9, N13 and N14, N16 and N17 and that a diffuse structured emission is present which contaminates the spectra (\citealt{Crivellari2009}). We thus use just the \chandra\ data for the spectral analysis (see section \ref{sec:spectrum}). We include also \xmm\ data, when possible, to study the point sources variability (see section \ref{sec:longvar}).

\subsection{Spectral analysis}
\label{sec:spectrum}

\begin{table*}
\begin{center} 
\begin{tabular}{ccccccccc} \hline
{Source} & {$n_{\rm H}$} & {${\it \Gamma}$} & {\em L$_{(0.5-10.0)\rm{ keV}}$} &{\em $\chi^{2}$/\rm{dof}} & {\em $P_{\rm val}$}  \\ 
 & 10$^{21} \rm {cm}^{-2}$ &   & 10$^{39}$ erg s$^{-1}$  & &\\
\hline
N9 & 4.1$^{+4.0}_{-2.8}$ &  1.9$_{-0.8}^{+0.9}$ & 17.8$\pm$1.8 & 0.48/3 & 0.92 \\

N10 & 5.3$_{-1.3}^{+1.7}$ & 1.9$_{-0.3}^{+0.3}$ & 76.8$\pm$3.4 & 24.0/22 & 0.35\\

N11 & 16.7$_{-7.8}^{+7.5}$ & 1.9$_{-0.7}^{+0.8}$ & 32.2$\pm$3.2  & 2.17/3 & 0.54\\

N14 &  3.1$_{-1.4}^{+1.8}$ & 2.7$_{-0.5}^{+0.6}$ & 15.1$\pm$1.3 & 6.8/5 & 0.23\\

N16 & $<$2.4 & 1.5$_{-0.4}^{+0.5}$ & 21.4$\pm$1.8 & 3.62/5 & 0.61\\

N17 & $<$1.6 & 2.1$_{-0.4}^{+0.5}$ & 11.5$\pm$1.0 & 1.38/5 & 0.93\\

\hline
\end{tabular}
\end{center}
\caption{ Average spectral shape of  the six sources with more than 100 counts in the sum of the three observations. The column density $n_{\rm H}$, the power-law index ${\it \Gamma}$, the unabsorbed luminosities in the (0.5-10.0) keV band, $\chi^{2}$/(degrees of freedom) and null hypothesis probability ($P_{\rm val}$) are listed. The model used is an absorbed power-law ({\sc tbabs*pow} in {\sc xspec}). Given the 2$\sigma$ (P$_{val}$ < 0.05) limit, all the fits are statistically acceptable.} 
\label{tab:free_unabs}
\end{table*}

\begin{table}
\begin{center} 
\begin{tabular}{ccccc} \hline
{ObsId} & {$n_{\rm H}$} & {${\it \Gamma}$} & {$\chi^{2}$/dof} & {$P_{val}$}\\
        & 10$^{21} \rm {cm}^{-2}$ & & & \\
\hline
2019 & 1.3$^{+0.7}_{-0.6}$ & 1.7$^{+0.2}_{-0.2}$ & \multirow{3}{*}{$\Bigg\}$ 41.32/56} &\\
9531 &  1.7$^{+2.0}_{-1.7}$ & 2.0$^{+0.3}_{-0.3}$ & & 0.93\\
9807 & $<$0.7 & 1.6$^{+0.2}_{-0.2}$ & & \\
all & 1.2$^{+0.5}_{-0.5}$ & 1.8$^{+0.2}_{-0.1}$ & 48.84/60 & 0.85\\

\hline
\end{tabular}
\end{center}
\caption{Spectral parameters for the average spectrum of all the sources (except N10) in each \chandra\ observation: the column density $n_{\rm H}$ and the power-law index ${\it \Gamma}$. The model used is an absorbed power-law ({\sc tbabs*pow} in {\sc xspec}). We leave all the parameters free to vary among the observations. 
In the last row, we report the result from the simultaneous spectral-fitting (all), with the parameters fixed among the three observations. 
We report also the $\chi^{2}$/dof and the null hypothesis probability ($P_{val}$) of the fit.}
\label{tab:av_spec}
\end{table}

Due to the relatively large distance of the Cartwheel group, most of the detected ULXs have low statistics and thus only the most luminous sources have enough counts to allow a spectral analysis. We thus attempt a spectral fit for the few brightest sources and we list the results in Table \ref{tab:free_unabs}.
We analyse the individual spectra for the 6 sources with at least 100 net counts, in the sum of the three \chandra\ observations (N9, N10, N11, N14, N16 and N17). We bin the data to have 20 total counts in each bin in order to apply the chi-squared statistics. 
To describe the spectral shape we use an absorbed power-law model, ({\sc tbabs*power-law} in {\sc xspec}), after background subtraction. 
The low statistics in the data does not allow us to use more complex models, such as two components models, which give a good description of the high-statistics ULXs spectra (e.g. \citealt{Gladstone2009b,Pintore2011}).  

For each source we combine the spectra obtained from the three images, using the {\sc combine\_spectra} procedure, to improve the statistics. The procedure sums multiple spectra combining also the associated background spectra and relative responses, and scaling background measures for both area and exposure time. In this way we can study the average spectral shape for each of these sources. 

We note that the spectra of the individual sources show a variety of slopes and different values of $n_{\rm H}$ absorption. In particular, we note that N11, the source with the largest $n_{\rm H}$, was already indicated in \citet{Wolter2004} as either a background source or an intrinsically very absorbed one due to the spectral counts distribution.

For the other sources there are not enough counts for studying the spectra. So we analyse an average \chandra\ spectrum combining data of all the sources, i.e. we extract a single spectrum adding together the counts of all the sources in each of the three observations, with {\sc specextract}, including also the sources fit individually, but excluding the most luminous ULX N10, which would bias the result due also to its large variability in the three epochs. 
The average spectrum in each epoch has been binned with 20 counts per bin and we use the chi-squared statistics in the fit. For the average spectrum, the background has been extracted from an elliptical region including all the galaxy but not the detected sources. 
From the simultaneous spectral fitting of the three \chandra\ epochs, with an absorbed power-law model, we obtain a total $n_{\rm H}$ = 1.2$^{+0.5}_{-0.5}\times$10$^{21}$ cm$^{-2}$ (including both the Galactic and intrinsic contribution) and ${\it \Gamma}$ = 1.8$^{+0.2}_{-0.1}$. The spectral shapes derived from the individual epochs are consistent with the values from the simultaneous fitting (see table \ref{tab:av_spec}). This suggests that no strong spectral changes are present in the average population. 
The Galactic absorption in the Cartwheel direction is $n_{\rm H}$ = 2.5$\times$10$^{20}$cm$^{-2}$, (\citealt{kalberla2005}).
We observe an additional local absorption, not surprising for ULXs. The presence of absorbing material in the vicinity of a source could be ascribed to the gas found in the Cartwheel ring \citep{Wolter2004} or to the powerful winds sometimes observed around ULXs (e.g. \citealt{middleton2015,pinto2016}).

We derive the flux of the low statistics sources by using a Bayesian
code ({\sc blike}\footnote{The library and the scripts used in section \ref{sec:longvar} can be found at
\url{https://github.com/andrea-belfiore/BLike}.}, A. Belfiore, in prep.). This computes a posterior
distribution for the expected source counts, starting from the
observed counts in the source and background regions, flat priors, and
assuming Poisson statistics. Then, assuming the spectral model in the last row of table \ref{tab:av_spec}, the instrument response and exposure time, we derive
estimates and upper limits on the flux of each source. In case of detections, we take the narrowest interval
covering 68 per cent of the posterior distribution and report
its central value as the flux estimator and its half-width as its uncertainty.
For consistency, we use the same spectrum and method to derive the fluxes in the single observations also for the six sources for which we have studied the spectrum. Even if N10 has not been included in the average spectrum, we give results for N10 with the same model. A detailed study of the flux, spectrum and variability of N10 is already published (\citealt{Wolter2006,Pizzolato2010}).  
The spectral model derived here gives fluxes consistent with those obtained in \citet{Wolter2004} (who used the spectral parameters $n_{\rm H}$ = 1.9$\times$10$^{21}$ cm$^{-2}$ and ${\it \Gamma}$ = 2.2): the difference on the flux is around 10 per cent.

In principle, a power-law may over-estimate the flux if a strong turn-over is present in the spectrum, as often observed in the ULX spectra (e.g. \citealt{Gladstone2009b}). We compare the fluxes obtained with the power-law model, with those obtained with a more complex model, i.e. a black body plus a multi-colour black body disc. We obtain the following best-fitting parameters: $n_{\rm H}$ = 1.6$\times$10$^{21}$ cm$^{-2}$, black body temperature T$_{bb}$ = 0.14 and disc temperature T$_{in}$ = 1.3, but the $n_{\rm H}$ is not well constrained ($n_{\rm H}$ $<$ 4$\times$10$^{21}$ cm$^{-2}$). The difference on the fluxes obtained from the one and the two components models is smaller than $\sim$ 25 per cent, in 2-10 keV, comparable with the average uncertainty on the flux.

We derive \xmm\ fluxes with the same spectral shape for comparison. The extraction radius used is 10 arcsec, except for N12 for which we use a smaller radius of 8 arcsec (the source is close to a CCD gap). We account also for the PSF encircled
fraction, which is 60/70 per cent in the adopted regions, applying a correction through the ARF file. The \xmm\ background is computed using an elliptical region which includes all the Cartwheel galaxy, excluding all the detected point sources.

The fluxes have been converted to luminosities assuming a distance of 131 Mpc. The unabsorbed luminosities of all the sources, derived with this method, are reported for each observation in Table \ref{tab:lmax_lmin}.

The 0.5-10 keV luminosity derived for the sources analysed individually, i.e. the luminosity in table \ref{tab:free_unabs}, is often larger than the average luminosity in table \ref{tab:lmax_lmin}. This difference is accounted for by the different spectral parameters, in particular the larger $n_{\rm H}$ derived in the individual fits (table \ref{tab:free_unabs}) with respect to the average model (last row of table \ref{tab:av_spec}).

\subsection{Short-term variability}
\label{sec:shortvar} 
Short-term variability $-$ within a single observation $-$ has been observed in ULXs, such as in 6 of 16 sources in archival \xmm\ observations analysed by \citet{Heil2009}, in Circinus ULX5 (\citealt{Walton2013,Mondal2021}) or in NGC 7456 ULX-1 \citep{Pintore2020}. In many cases, \citet{Heil2009} find less than the variability expected assuming a similar behaviour to Galactic BH binaries (GBHBs) or Active Galactic Nuclei (AGN). They conclude that this apparent lack of variability is intrinsic to the sources, not a consequence of insufficient observations. They propose that, if the ULXs variability is similar to that of GBHBs, only a ULX spectral component similar to the thermal component associated to the accretion disc of GBHBs, with little short-term variability, is visible in the \xmm\ band-pass. Otherwise, there must be a mechanism which suppresses variability in ULXs or an accretion process with very stable X-ray emission, not observed in GBHBs and AGNs. Despite not finding an obvious correlation between variability and luminosity, their analysis suggests that many brighter ULXs have suppressed short-term variability.
Considering the distance of the Cartwheel and the consequent low statistics of our data, it is sensible to search for short-term variability (time-scales of minutes-hours) only in the brightest sources. Furthermore, taking into account these previous results on variability of bright sources, we do not expect to find a wide occurrence of short-term variability. 

In order to investigate the variability of the sources in a single observation we select all the sources with at least 20 total counts, in 0.5-10 keV, in the single \chandra\ observation. We check that the background does not give a contribution to the variability. A flare in the background emission is present just in observation 9531 and has been excluded (see section \ref{sec:datared}). 
We apply a Kolmogorov-Smirnov test on the arrival time of the photons comparing it with a uniform expected distribution: 
if there is no variability, we expect the photons to arrive at regular time intervals on the telescope.

The analysis has a negative result; only P26 in observation 9807 shows variability within the observation at $\sim 3 \sigma$.

We do not investigate the issue further.

\subsection{Long-term variability}
\label{sec:longvar}
We study the long-term variability in the (0.5-10 keV) band across the three \chandra\ observations (table \ref{chandra_xmm_obs}) and, for resolved sources, including also \xmm\ data.  
We assess and measure this variability in three, somewhat complementary, ways:
first we single out the variable sources by fitting the observed counts to
a constant luminosity model; for those variable at $\geq$ 3$\sigma$, we measure
the largest luminosity ratio ($L_{\rm max}/L_{\rm min}$) between different observations and estimate
the relative excess variance in their light curve.
All these tests are carried out with some care due to the very low counts.

We start from model fitting because it applies to all light curves and it is a robust and sound preliminary test. If a constant model explains the observed data well enough, then the source variability is not worth investigating further. The luminosity ratio quantifies in absolute terms the minimum range of luminosities observed. The fractional variability describes the average behaviour of the source, not only the extreme values. Because it depends on the error bars it  characterises better bright sources.

We derive, with {\sc blike}, the posterior distribution for the expected net source counts, in each observation, starting from the counts observed in the source and background regions. 
We assume Poisson statistics and flat priors for both source and background counts and rescale the background contribution proportionally to the areas of the two regions.  In order to convert from counts to flux, and thence to luminosity,
we consider the duration $T$ of each observation, the source spectrum convolved
with the response of the instrument at each epoch (to take into account the decrease over the years of the \chandra\ effective area) and the encircled fraction.
We work out this conversion factor using {\sc xspec}, approximating all
and the last row of table \ref{tab:av_spec}). 
We assume as a null hypothesis that the luminosity, $L_{\rm ave}$ in table \ref{tab:lmax_lmin}, does not change between observations. We obtain this value by maximising the combined likelihood $\mathcal{L}_{con}$ over the average common luminosity. 
The alternative hypothesis of a variable source maximises the likelihood in each single observation to a different luminosity and combines these values of the likelihood into $\mathcal{L}_{var}$.
We measure the test statistics $TS = -2 * log(\mathcal{L}_{con}/\mathcal{L}_{var})$. For a constant source, according to Wilks' theorem \citep{Wilks1938}, $TS$ distributes as
a $\chi^{2}_{n}$ with $n=N_{\rm obs} - 1$ degrees of freedom, where $N_{\rm obs}$ is the number of observations. We use this
asymptotic behaviour to assess the variability of each source: we compare the $TS$ to a $\chi^{2}_{n}$ distribution and convert the $P_{value}$ into sigma units (reported in column 10 of table \ref{tab:lmax_lmin}). We consider as variable a source where a constant luminosity can be rejected beyond 3$\sigma$. 
We evaluate the 1-$\sigma$ error bars from the likelihood profile around $L_{\rm ave}$: in this case $TS$ distributes as a $\chi^{2}_{1}$.
This approach allows us to fit also observations where the source is not
detected.

We extend our analysis to the sources resolved by \xmm\, i.e. N6, N10, N11, N12, N21, P25, 
with the extraction regions and the spectra as described in section \ref{sec:spectrum}. The expected difference in flux calibration between \chandra\ and \xmm\ is $\sim$ 15 per cent \citep{Madsen2017}, usually smaller than the average uncertainty on the flux.

For each variable source we measure the maximum range of variability by computing
the luminosity ratio $L_{\rm max}/L_{\rm min}$ between different observations.
This estimate and its uncertainty are obtained by bootstrapping from the posterior
luminosity distributions for any two pairs of observations. We then extract the
smallest 68 per cent coverage interval for the luminosity ratio and take the middle
point as our estimate and its half amplitude as its uncertainty.
Finally we take the largest value of this ratio between any two pairs of
observations and report it in Table \ref{tab:lmax_lmin}, column 8.

A common method to quantify the average variability (e.g. \citealt{Sutton2013}) relies on the fractional variability amplitude $F_{\rm var}$ (e.g. \citealt{edelson2002,Vaughan2003}) derived from the excess variance \citep{Nandra1997}. We slightly modify the original formula to take into account the upper limits:
\begin{equation}
     F_{\rm var} = \frac{1}{L_{\rm ave}}\sqrt{S^{2}-<\sigma_{\rm err}^{2}>}
    \label{eq:Fvar}
\end{equation}
where $S^{2}$ is the total variance of the light curve (assuming no errors), $<\sigma_{\rm err}^{2}>$ is the mean error squared. The values and errors are defined as the mid-point and half-width of the highest density interval for the luminosity in each observation. In case of non-detection both value and error coincide with half of the 68 per cent upper limit. The light curve variance is defined as: 
\begin{equation}
  S^{2} = \frac{1}{N_{\rm obs}-1}\displaystyle\sum_{i=1}^{N_{\rm obs}} \left(L_{i} - \Bar{L}  \right)^{2}
\end{equation}
with $L_{i}$ the luminosity in the observation $i$ and $\Bar{L} = \displaystyle\sum_{i=1}^{N_{\rm obs}} \dfrac{L_{i}}{N_{\rm obs}}$ is the arithmetic mean. 
The uncertainty in $F_{\rm var}$ is:
\begin{equation}
   F_{\rm var}^{\rm err} = \sqrt{\left(\sqrt{\frac{1}{2N_{\rm obs}}} \frac{<\sigma_{\rm err}^{2}>}{L_{\rm ave}^{2}{F_{\rm var}}} \right)^{2} + \left(\sqrt{\frac{<\sigma_{\rm err}^{2}>}{N_{\rm obs}}} \frac{1}{L_{\rm ave}} \right)^{2}}
\end{equation}

These quantities are reported in table \ref{tab:lmax_lmin}, column 9, and discussed in section \ref{sec:longterm_var_res}.
\begin{table*}
\begin{center} 
\scalebox{0.9}{%
\begin{tabular}{cccccccccc} \hline
{Source} & $L_{2019}$ & 
$L_{101}$ & $L_{201}$
& $L_{9531}$ &  $L_{9807}$ & $L_{\rm ave}$ & $\dfrac{L_{\rm max}}{L_{\rm min}}$ & {\em F$_{\rm var}$} & {$\sigma$}\\ 
(1) & (2) & (3) & (4) & (5) & (6) & (7) & (8) & (9) & (10)\\
\hline
N2 (G1) & 7.6$\pm$1.3 &  &  & 7.7$\pm$1.7 & 5.2$\pm$1.4 & 7.0$\pm$0.8 &  &  & 0.8 \\
N3 (G1) & 11.2$\pm$1.5 &  &  & 25.6$\pm$3.1 & 18.4$\pm$2.6 & 16.9$\pm$1.2 & 2.2$\pm$0.4 & 0.40$\pm$0.09 & \textbf{4.2} \\
N4 (G1) & 9.1$\pm$1.4 &  &  & 5.8$\pm$1.5 & 7.2$\pm$1.7 & 7.6$\pm$0.8 &  &  & 1.1 \\
N5 (G1) & 7.3$\pm$1.3 &  &  & $<$2.7 & 2.9$\pm$1.1 & 4.7$\pm$0.7 & 3.1$\pm$1.5 & 0.55$\pm$0.14 & \textbf{3.1} \\
N6 & 3.4$\pm$0.8 & $<$1.7 & $<$5.8 & $<$1.7 & $<$1.8 & 2.2$\pm$0.5/2.2$\pm$0.4 &  &  & 1.9/1.9 \\
N7 & 12.4$\pm$1.6 &  &  & 3.3$\pm$1.1 & 2.4$\pm$1.1 & 7.5$\pm$0.8 & 4.6$\pm$2.1 & 0.73$\pm$0.10 & \textbf{5.3} \\
N9 & 11.6$\pm$1.5 &  &  & 18.0$\pm$2.6 & 9.2$\pm$1.9 & 12.7$\pm$1.1 &  &  & 2.5 \\
N10 & 68.4$\pm$3.7 & 85.7$\pm$7.0 & 50.0$\pm$4.5 & 53.7$\pm$4.4 & 17.5$\pm$2.6 & 51.7$\pm$2.2/54.5$\pm$1.5 & 3.8$\pm$0.6/4.8$\pm$0.8 & 0.50$\pm$0.04/0.45$\pm$0.04 & \textbf{10.0}/\textbf{10.8} \\
N11 & 12.4$\pm$1.6 & $<$8.0 & $<$5.2 & 12.1$\pm$2.1 & 9.6$\pm$1.9 & 11.5$\pm$1.0/10.5$\pm$0.8 &  &  & 0.7/1.9 \\
N12 & 7.9$\pm$1.3 & $<$2.7 & $<$5.8 & 7.7$\pm$1.7 & 3.4$\pm$1.2 & 6.7$\pm$0.8/6.2$\pm$0.7 &  &  & 2.1/2.5 \\
N13 & 7.0$\pm$1.2 &  &  & 8.8$\pm$1.8 & 9.7$\pm$2.0 & 8.1$\pm$0.8 &  &  & 0.7 \\
N14 & 15.4$\pm$1.8 &  &  & 14.6$\pm$2.3 & 18.5$\pm$2.7 & 15.9$\pm$1.1 &  &  & 0.7 \\
N15 & 3.1$\pm$0.8 &  &  & 3.6$\pm$1.2 & 2.2$\pm$0.9 & 2.9$\pm$0.5 &  &  & 0.4 \\
N16 & 18.8$\pm$2.0 &  &  & 12.3$\pm$2.1 & 18.1$\pm$2.6 & 16.7$\pm$1.2 &  &  & 1.7 \\
N17 & 20.5$\pm$2.1 &  &  & 14.1$\pm$2.3 & 6.3$\pm$1.6 & 15.1$\pm$1.2 & 3.1$\pm$0.8 & 0.46$\pm$0.08 & \textbf{4.7} \\
N18 & 1.3$\pm$0.5 &  &  & $<$1.3 & $<$1.1 & 1.1$\pm$0.5 &  &  & 0.5 \\
N19 (G1) & 2.5$\pm$0.8 &  &  & 5.2$\pm$1.4 & 2.7$\pm$1.0 & 3.3$\pm$0.6 &  &  &  1.4\\
N20 & 1.7$\pm$0.6 &  &  & $<$0.4 & $<$0.4 & 0.8$\pm$0.3 & 7.0$\pm$6.4 & 1.09$\pm$0.33 & \textbf{3.0} \\
N21 & 1.3$\pm$0.6 & $<$1.4 & $<$4.5 & 6.0$\pm$1.5 & 2.3$\pm$1.0 & 2.7$\pm$0.5/2.5$\pm$0.5 &  &  & 2.9/2.9 \\
N22 & 2.7$\pm$0.8 &  &  & 3.0$\pm$1.1 & 2.6$\pm$1.1 & 2.7$\pm$0.5 &  &  & 0.0 \\
N23 & 2.1$\pm$0.7 &  &  & $<$0.4 & $<$0.7 & 0.9$\pm$0.3 &  &  & 2.8 \\
N24 & $<$1.6 &  &  & $<$1.5 & $<$2.0 & 1.2$\pm$0.4 &  &  & --\\
P25 & $<$0.7 & $<$3.7 & $<$5.6 & 10.4$\pm$2.0 & 8.1$\pm$1.7 & 5.0$\pm$0.7/4.8$\pm$0.6 & 18.7$\pm$12.8/18.5$\pm$12.8 & 1.01$\pm$0.18/0.80$\pm$0.18 & \textbf{6.7}/\textbf{6.3} \\
P26 & $<$0.7 &  &  & $<$1.3 & 11.7$\pm$2.2 & 3.0$\pm$0.6 & 20.8$\pm$14.9 & 2.06$\pm$0.25 & \textbf{6.9} \\
P27 & 4.8$\pm$1.0 &  &  & 4.8$\pm$1.3 & $<$2.6 & 4.0$\pm$0.6 &  &  & 1.6 \\
P28 & $<$0.7 &  &  & 2.9$\pm$1.1 & 2.6$\pm$1.0 & 1.6$\pm$0.4 &  &  & 2.3 \\
P29 & $<$1.2 &  &  & $<$1.3 & 3.1$\pm$1.1 & 1.4$\pm$0.4 &  &  & 1.8 \\
P30 (G1) & $<$0.5 &  &  & 5.4$\pm$1.5 & $<$1.4 & 1.6$\pm$0.4 & 13.2$\pm$10.6 & 1.75$\pm$0.37 & \textbf{4.2} \\
P31 (G1) & 6.2$\pm$1.2 &  &  & 5.5$\pm$1.4 & 5.7$\pm$1.5 & 7.4$\pm$0.8 &  &  & 0.1 \\
P32 (G3)$^{*}$ & 1.8$\pm$0.6 &  &  & 2.9$\pm$1.1 & 3.0$\pm$1.2 & 2.3$\pm$0.5 &  &  & 0.5 \\
P33 & $<$0.5 &  &  & 2.3$\pm$1.0 & $<$1.3 & 0.8$\pm$0.3 &  &  & 2.0 \\
P34 & $<$0.2 &  &  & $<$0.4 & 3.5$\pm$1.2 & 0.5$\pm$0.3 & 25.4$\pm$23.1 & 3.58$\pm$0.85 & \textbf{4.2} \\
P35 & $<$1.1 &  &  & 2.5$\pm$1.0 & $<$1.9 & 1.3$\pm$0.4 &  &  & 1.0 \\
P36 & 1.9$\pm$0.7 &  &  & $<$2.3 & $<$0.8 & 1.3$\pm$0.4 &  &  & 1.5 \\
P37 & 1.9$\pm$0.6 &  &  & $<$1.5 & $<$0.6 & 1.2$\pm$0.3 &  &  & 1.8 \\
P38 & 2.1$\pm$0.7 &  &  & $<$1.5 & 4.0$\pm$1.3 & 2.2$\pm$0.5 &  &  & 1.6 \\
P39 & $<$0.2 &  &  & 2.5$\pm$1.0 & $<$0.7 & 0.5$\pm$0.3 & 16.3$\pm$15.1 & 2.18$\pm$0.73 & \textbf{3.1} \\
P40 & $<$0.5 &  &  & $<$0.7 & 2.6$\pm$1.0 & 0.7$\pm$0.3 &  &  & 2.5 \\
P41 & $<$0.4 &  &  & $<$2.1 & $<$2.0 & 0.6$\pm$0.3 &  &  & --\\
P42 (G1n)  & $<$1.4 &  &  & $<$0.7 & $<$1.4 & 0.7$\pm$0.3 &  &  & --\\
P43 (G2n) & 1.4$\pm$0.6 &  &  & 4.2$\pm$1.3 & 2.5$\pm$1.0 & 2.4$\pm$0.5 &  &  & --\\
P44 (G3n) & 2.3$\pm$0.8 &  &  & $<$2.1 & $<$3.1 & 2.0$\pm$0.5 &  &  & --\\

\hline
\end{tabular}
}
\end{center}
\caption{Column 1: names of the sources detected in the Cartwheel group: we use the same nomenclature of Table 1 in \citet{Wolter2004} for
sources that we have matched in common with that study and "P", with progressive numbers, for sources not listed in Table 1 of \citet{Wolter2004}; Columns 2, 5, 6: unabsorbed luminosity (or 1$\sigma$ upper limit) of the ULXs in the \chandra\ observations; Columns 3, 4: unabsorbed luminosity (or 1$\sigma$ upper limit) of the ULXs in \xmm\ observations, all the luminosities (in units of $10^{39}$ erg s$^{-1}$) are computed in (0.5 - 10) keV band, and have been derived assuming a distance of 131 Mpc and an absorbed power-law ($n_{\rm H}$ = 1.2$\times$10$^{21}$ cm$^{-2}$ and ${\it \Gamma}$ = 1.8; see \ref{sec:spectrum}); Column 7: unabsorbed average luminosity. We report also, as an estimation of the variability, Column 8: the maximum luminosity ratio (see text for more details on the computation); Column 9: the fractional variability; Column 10: variability significance from the fit to a constant flux; we highlight in bold the significance values $\geq$3$\sigma$, which we use as a threshold to determine the variable sources. Sources N24 and P41 are excluded from the variability analysis because they have no significant flux detections and P42, P43, P44 are excluded because they do not classify as ULXs. 
$^{*}$ P32 was already present in the analysis of \citet{Wolter2004}, but not indicated as N because it was out of the field used in their table 1. \\
When two values are listed in the table, the value on the left refers to \chandra\ data only, the value on the right considers both \chandra\ and \xmm\ data. }
\label{tab:lmax_lmin}
\end{table*}

\subsection{X-ray Luminosity function}
\label{sec:lum_func}
The luminosity function (LF) of a population of sources describes how they distribute as a function of their luminosity. Therefore it can be interpreted as an instantaneous picture of the population considered.
Having three \chandra\ epochs spanning eight years, we want to verify if the Cartwheel XLF remains consistent with a single shape, despite the long-term variability of its sources.
\citeauthor{Zezas2006} (\citeyear{Zezas2006,Zezas2007}) study the variability of point sources in the Antennae and conclude that, although most sources exhibit flux variability of a factor 2-6 in months to years time-scales, the total XLF remains consistent with a single shape.
The Cartwheel cumulative XLF of the first epoch has already been computed in \citet{Wolter2004}, where the authors find that the XLF of the first \chandra\ observation (considering only the sources in the ring) agrees with the "universal" LF for High Mass X-ray Binaries (HMXBs) \citep{Grimm2003}, assuming a Star Formation Rate (SFR) of 20 $M_{\odot}$ yr$^{-1}$. The authors observe also a flattening of the XLF at low luminosity due to detection incompleteness, but they do not correct the distribution. 
The background emission is not constant across the Cartwheel galaxy, e.g. a larger background emission is observed in the ring, thus the source detection is hindered in the regions with the largest background emission. 
We produce an incompleteness-corrected XLF for the Cartwheel galaxy following the same procedure used e.g. in \citet{Wolter2015}: 
we use figure 12 from \citet{kim2003} in which the detection probabilities are plotted as a function of background counts and source counts. We refer to the second panel, off-axis of 2 arcmin,
since the Cartwheel sources are always within this off-axis limit, interpolating linearly between counts lines. The background level is derived for each source individually, by using the background annular regions described in appendix \ref{appendixA} and it is included in the range 0.03-0.18 cts pixel$^{-1}$. For each of the three epochs, we consider all the ULXs which are associated with the Cartwheel (while only ring sources are used in \citealt{wolter2018}) and with a flux detection with significance $\geq$ 3$\sigma$ according with {\sc blike}. 
We compute the cumulative XLF 
in the (2.0-10.0) keV band for consistency with previous works and to compare with the \citet{Grimm2003} function. The (2-10) keV luminosities are reported in table \ref{tab:netcont} in appendix \ref{appendixA}. The choice of the hard band also reduces the possibility of contamination of a hot diffuse plasma from the galaxy or the compact group to which the Cartwheel belongs i.e. the background emission, which peaks in the soft band, and of a possible local absorption.

\begin{figure*}
\begin{center}

        \includegraphics[scale=0.4]{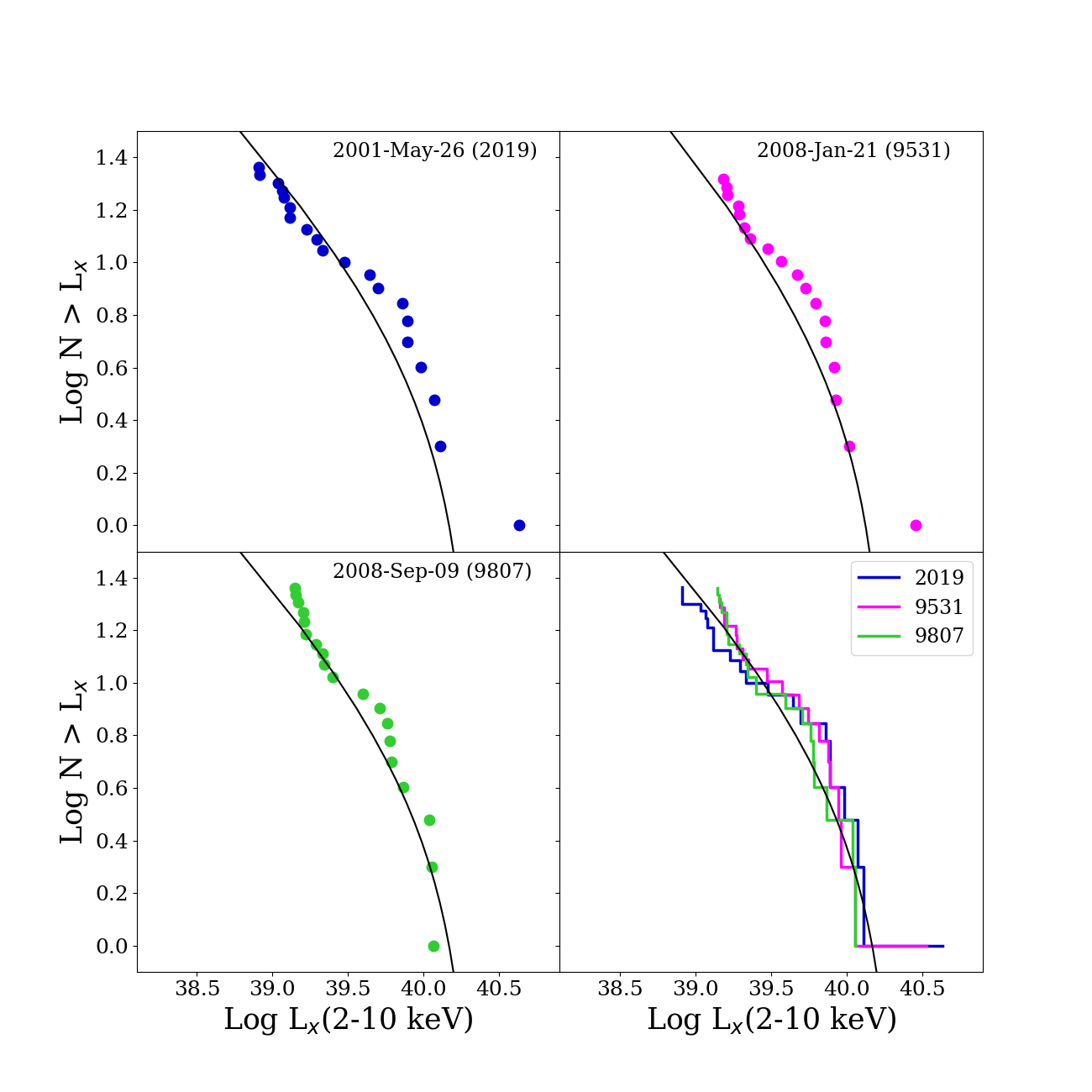}

        \caption{Cumulative XLF of the Cartwheel galaxy with \chandra\ data (upper panels and bottom left). In the bottom right panel the three \chandra\ observations are plotted together. The black line is the \citet{Grimm2003} function for HMXBs, assuming a SFR of 20 $M_{\odot}$yr$^{-1}$, consistent with various measures for the Cartwheel galaxy (see text).}\label{fig:LumFunct}
\end{center}
\end{figure*}

In all the epochs, the cumulative XLF, plotted in Figure \ref{fig:LumFunct}, remains consistent with the \citet{Grimm2003} function assuming 20 $M_{\odot}$ yr$^{-1}$ SFR, as previously found by \citet{Wolter2004} for the first \chandra\ observation. 
In addition, the XLF does not have large differences in shape in the three epochs, considering also that there is an average error on the luminosity of $\sim$ 25 per cent, except for the most luminous source.
Therefore we can construct the differential XLF summing the ULXs of the three observations to increase the statistics.  
The observations of year 2008 have a higher detection limit, that corresponds to a (2-10) keV luminosity of $\sim$1.4$\times$10$^{39}$ erg s$^{-1}$, with respect to the first one. We bin the sources starting from this luminosity in order to exclude a lower bin with a large incompleteness where only observation 2019 would contribute. This results in 49 detections, counting all sources in all the different observations. The luminosities used are underlined in Table \ref{tab:netcont} column 3, 5, 7 in appendix \ref{appendixA}. We also correct for the loss of sources due to high diffuse background, following \citet{kim2003}, as done for the cumulative XLF, thus renormalizing the density.   
We group the sources in five equally logarithmic spaced bins and  divide the number of sources in each bin by the width of the bin.

We fit the differential LF with a power-law model, using a maximum likelihood approach. We assume that the luminosities are distributed as $f$=$n\, L_{39}^{-\gamma}$, where $n$ is the normalization, $\gamma$ the power-law index and $L_{39}$ is the luminosity in units of 10$^{39}$ erg s$^{-1}$. 
The detailed procedure for deriving the power-law likelihood is given in the appendix \ref{appendixB}.
The best fit is the one that maximises the log likelihood function.

We evaluate the goodness of fit applying a Kolmogorov-Smirnov test, which compares the empirical cumulative distribution function (CDF) of the observed sample with the theoretical distribution, i.e. power-law.  
In the evaluation of the CDF, binning is not needed.
The result of the fit to the model is reported in Table \ref{tab:fit_diff_XLF}: the Pvalue is larger than 5 per cent, which we choose as a threshold for the probability, thus we cannot reject the null hypothesis, i.e. the observed sample is drawn from the assumed distribution.
\newline
\newline
\begin{figure*}
\begin{center}

      \includegraphics[scale=0.45]{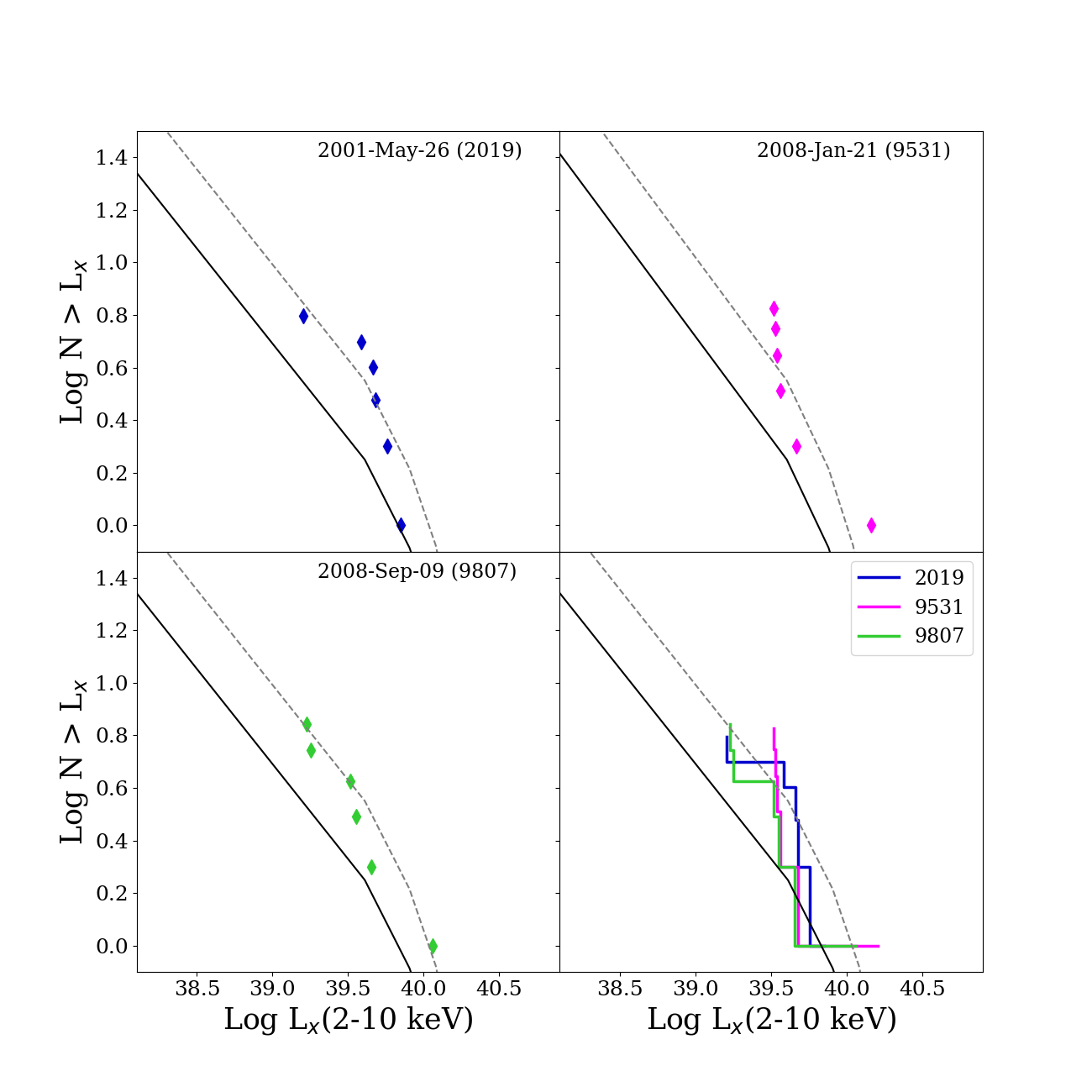}
        
        \caption{Cumulative XLF of the galaxy G1, with \chandra\ data (upper panels and bottom left). In the bottom right panel the three \chandra\ observations are plotted together. The solid-black line and the dashed-grey line are the \citet{Grimm2003} function for HMXBs, assuming respectively a SFR of 5 and 10 $M_{\odot}$yr$^{-1}$. }
        \label{fig:CumXLFG1}
\end{center}
\end{figure*}

For completeness, we construct the cumulative XLF for the galaxy G1 (see figure \ref{fig:CumXLFG1}; the luminosities used are reported in table \ref{tab:netcont} and highlighted in italic). Its shape is more similar to the \citet{Grimm2003} function for HMXBs, with a SFR of $\sim$ 10 $M_{\odot}$yr$^{-1}$, suggesting a larger SFR than the value derived by \citet{Crivellari2009} of 5 $M_{\odot}$yr$^{-1}$.

\begin{table}
\begin{center} 
\begin{tabular}{cccc} \hline
{\em $\gamma$} & {\em $n$} & {\em P$_{val}$}\\
\hline
1.76$^{+0.51}_{-0.43}$ & 59$^{+51}_{-28}$ & 0.51\\
\hline
\end{tabular}
\end{center}
\caption{Parameters obtained from the fit of the Cartwheel X-ray differential luminosity function, with a power-law. $\gamma$ is the power-law index; $n$ the normalization; P$_{val}$ is the null hypothesis probability.  
The uncertainties are given at 1$\sigma$ level.}
\label{tab:fit_diff_XLF}
\end{table}

\begin{figure}
\begin{center}
      \includegraphics[scale=0.45]{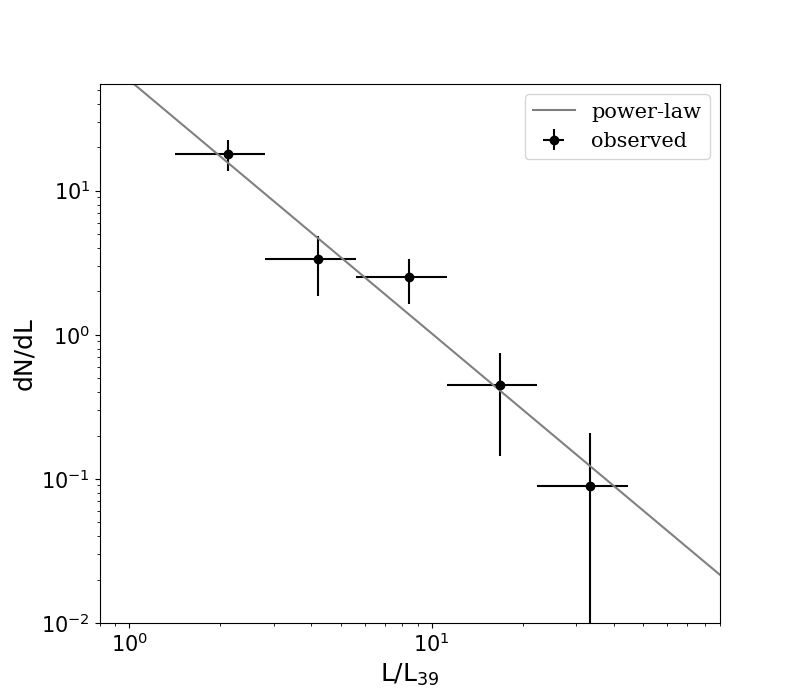}
        
        \caption{Differential XLF of the Cartwheel galaxy. The observed values are the black points and the Y-axis errors are statistical uncertainties in Gehrels approximation \citep{Gehrels1986}. The statistical uncertainty in the bin at highest luminosity is large because it contains just two ULX, thus the lower error bar extends below the lower dN/dL value displayed in the plot. The X-axis error bars indicates the width of each bin. The grey line is the power-law model with $\Gamma$ = 1.76$_{-0.43}^{+0.51}$. }
        \label{fig:plotfitdiffXLF}
\end{center}
\end{figure}

\subsection{Diffuse emission}
\label{sec:diff}
Both the Cartwheel and its companions show diffuse emission in the X-ray range (\citealt{Wolter2004,Crivellari2009}). In principle this could be due to both an extended plasma around or in the galaxies, or to non-resolved X-ray binaries. If the latter, it could be itself variable and therefore affect the variability and spectral properties of the analysed sources. We thus analyse this component. No variability is detected in the diffuse component within the uncertainties, suggesting that it does not affect the point sources  variability. We report the analysis of the diffuse component in appendix \ref{appendixC} for completeness.

\section{Results}
\label{sec:results}
\subsection{Point sources}
\label{sec:poinsrc-res}
We detect in total 44 different individual sources, at least in one observation or in their combination. 
We compare the results for \chandra\ observation 2019 of year 2001 with \citet{Wolter2004}. Table 1 in \citet{Wolter2004} contains 24 sources in the field of view of Carthweel, G1 and G2 only (G3 is outside the chosen area but still inside the \chandra\ field of view). In this instance we recover a somewhat different set of sources, due to slightly different parameters used. Source N18 is not recovered, while sources P31 and P37 are included (these last ones were listed in \citealt{gao2003}). These effects are due to low statistics and size of the detection kernel. 
Source P32, associated with G3, was listed in Appendix A in \citet{Wolter2004}. 
Ten more sources are detected in either or both the \chandra\ observations of year 2008, due to flux variability. 
Furthermore, by exploiting the merged image we reach deeper fluxes, which allow us to detect 8 further sources, see table \ref{tab:netcont} in appendix \ref{appendixA}. One of these (P36) was in the list of \citealt{gao2003}. 
Among the "new" \chandra\ sources we find also the \xmm\ source XMM8 \citep{Crivellari2009} both in the \chandra\ observations of 2008 and in the merged image, and we name it P25 for consistency. In total we find 15 
new sources, not published previously (we mark them with the simbol $^{\dagger}$ in table \ref{tab:netcont} in appendix \ref{appendixA}).

We exclude sources N1 and N8 rom the variability analysis and characterisation of the XLF, because not positionally associated to the group galaxies. 
Three of the newly detected sources, P42, P43 and P44, are probably linked to the nuclear region of G1, G2 and G3 (see table \ref{tab:g1_g2_g3}). Therefore they do not qualify as ULXs and we exclude them as well. Their luminosity is $\sim$ 10$^{39} $ erg s$^{-1}$ indicating that no bright AGN is present in any of these galaxies.  
N24 and P41, have no flux detection above 3$\sigma$, but just upper limits, therefore we do not consider them for the variability analysis. 
In summary, the total number of sources positionally associated to the group galaxies, for which we can investigate variability, are 37 in total: Cartwheel has 29 ULXs; G1 has 7 ULXs (plus 1 possibly nuclear source); G2 has only 1 possibly nuclear source; G3 has 1 ULX (plus 1 possibly nuclear source).

\subsection{Intensity variability}
\label{sec:longterm_var_res}

Fluxes of ULXs are often found to be variable on different time-scales (e.g. \citealt{Heil2009,Earnshaw2018}). We have analysed both the short-term variability (minutes-hours) in the individual \chandra\ observations and the long-term (months-years) variability among the observations including \textit{XMM-Newton} data, when possible, for the ULXs in the Cartwheel and in its companion galaxies.

The short-term variability, usually poorly predictable in ULXs because it is not necessarily present in all the observations of the same source (e.g. \citealt{Sutton2013}), is rarely found and at low significance. 
The small statistics of the available observations does not compel to elaborate further.

Instead, the long-term variability is frequent in the light curves analysed here: out of the 37 sources in the Cartwheel group, we count 11 sources variable at more than 3$\sigma$ significance: 3/7 in G1 and 8/29 in the Cartwheel galaxy.  

Among the variable sources, N20, P25, P26, P30, P34 and P39 are transients i.e. detected at least once above and once below 10$^{39}$ erg s$^{-1}$, or once above and once with an upper limit $\leq$ 10$^{39}$ erg s$^{-1}$.  Since a detection or a non-detection can depend also on the quality of the observations, low-statistics short observations have to be treated carefully. Possibly, sources N23, P28, P33, P35, P36, P37 and P40 could satisfy the definition of transients, with more statistically stringent measures. 

The minimum luminosity ratio $L_{\rm max}$/$L_{\rm min}$ observed among the variable ULXs in 
the Cartwheel group is $\sim$2.2 (source N3), while 5 of the 6 transient sources  
have a luminosity ratio of about one order of magnitude. The $F_{\rm var}$ statistics measures the average intensity variability: $\sim$ 70 per cent of the variable sources (8/11) have $F_{\rm var}$ larger than 50 per cent (table \ref{tab:lmax_lmin}).

\begin{figure*}
\begin{center}

        \includegraphics[scale=0.30]{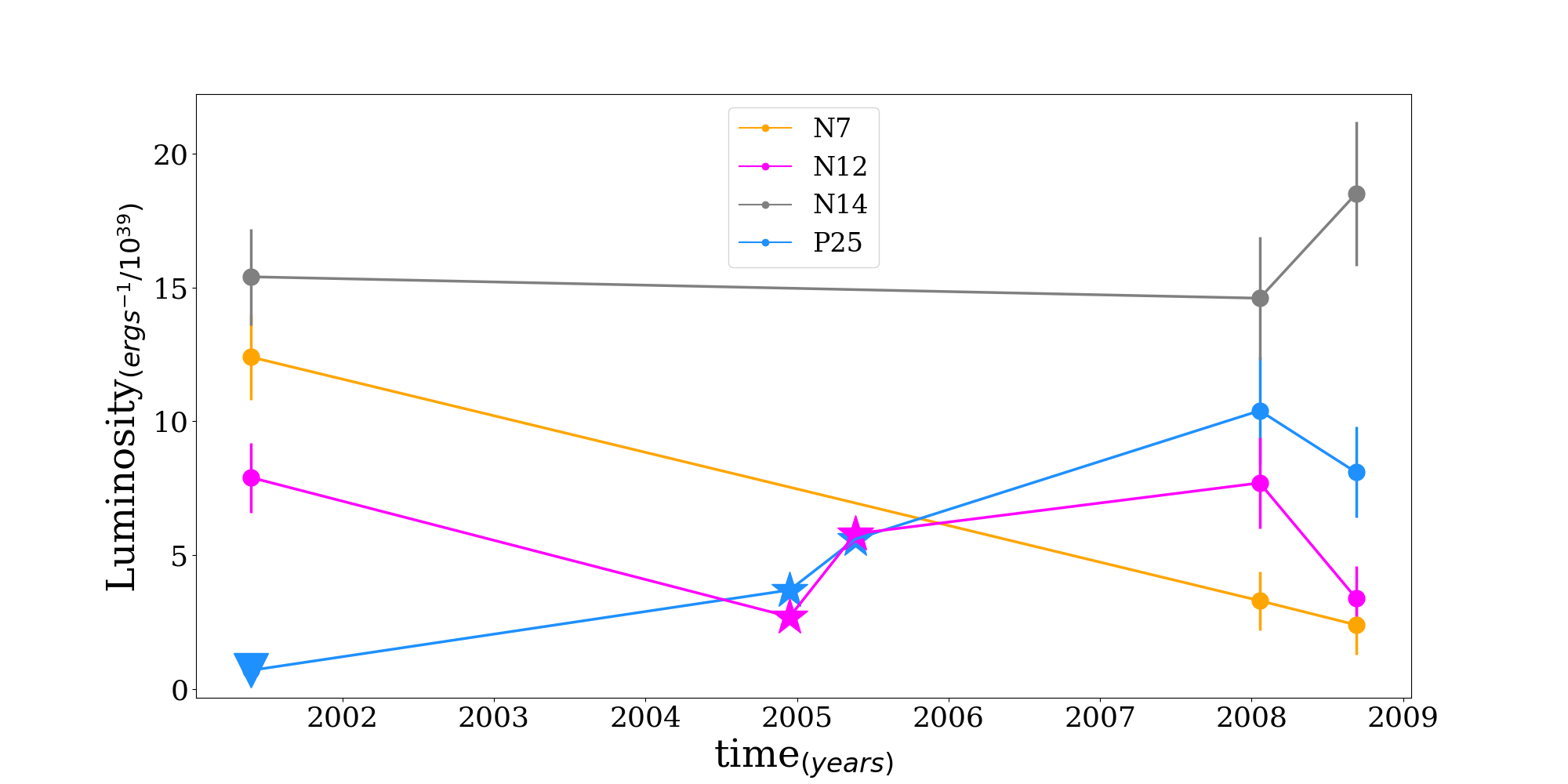}

        \caption{The light curves of 4 sources in the Cartwheel galaxy: N7, N12, N14, P25. 'Circles' indicate the \chandra\ observations, 'stars' the \xmm\ 1$\sigma$ upper limits, 'inverted triangles' are the 1$\sigma$ \chandra\ upper limits, derived with the code {\sc blike}.}\label{fig:lc}
\end{center}
\end{figure*} 

To show the observed variability  we plot in 
Figure~\ref{fig:lc} a few representative light curves: P25 as an example of a transient source, N7 with luminosities near the instrumental detection threshold in the second and third \chandra\ exposures; N12 with upper limits larger than 10$^{39}$ erg s$^{-1}$, thus not classified as transient; and N14, a luminous source well above the detection threshold of the instrument.

\subsection{X-ray Luminosity Function}
\label{sec:XLF_results}
We model the differential XLF of the Cartwheel with a power-law model. We obtain a slope $\gamma$ = 1.76$_{-0.43}^{+0.51}$. 
The result is consistent within the uncertainties with the one reported by \citet{Grimm2003} for HMXBs ($\gamma = 1.61 \pm 0.12$). 
\citet{swartz2011} find $\gamma$ = 1.4$\pm$0.2, with a power-law, consistent with our fit within the uncertainties.

More recently \citet{Lehmer2019} run a similar analysis for a large sample of SINGS galaxies with \chandra\ observations, disentangling the contributions of LMXB and HMXB for galaxy subsamples. Their overall fit for HMXBs has a slope consistent with our result, but with an exponential cut-off at luminosity $L_{c}$ = 5.0$^{+7.6}_{-1.9} \times$10$^{40}$ erg s$^{-1}$. 
We fit our XLF with an exponential cut-off power-law model ($f=nL^{-\gamma}_{39}\exp(-L_{39}/L_{c,39}$), where $L_{c,39}$ is the cut-off luminosity in units of 10$^{39}$ erg s$^{-1}$), in order to find the cut-off for our sample, but we do not find the presence of a cut-off, possibly due to the low statistics of our sample or to the presence of higher luminosity sources.

We can use the relationships for the XLF (e.g. \citet{Grimm2003}:  $dN/dL_{38}$ = 3.3SFR $L^{-1.61}$ for $L<$2.1$\times$10$^{40}$ erg s$^{-1}$)  also to derive the expected SFR. Fitting our XLF, excluding sources above 2.1$\times$10$^{40}$ erg s$^{-1}$, fixing $\gamma = 1.61$ and comparing the resulting normalisation with 3.3SFR, we obtain an estimate of the Cartwheel SFR of 20.7 $M_{\odot}$ y$^{-1}$. This value is not far from the more recent values found in the literature: 17.7 $M_{\odot}$ y$^{-1}$ and 18 $M_{\odot}$ y$^{-1}$, derived respectively from FIR data and $L_{H_{\alpha}}$, see \citet{mayya2005}.

\section{Discussion}
\label{sec:discussion}
In the Cartwheel galaxy and companions, the variable ULXs are 11/37 ($\sim$ 30 per cent) of which 6 are transients. This fraction should be regarded as a lower limit, since the number or observations is small and we do not know the duty cycle of these sources, and because sources at the detection limit might not be classified as variable just due to poor statistics. Three of the variable sources are in G1, out of 7 detected ULXs, which gives an higher fraction of variable sources of $\sim$ 43 per cent. We can compare our results to those for a similar sample of variable sources studied in the Antennae galaxies \citep{Zezas2006}. The Antennae ULXs, which are the sample with the highest statistics in the datasets used, and far from the detection limits, are 14, observed over 7 different epochs. Of these, 11 (78 per cent), of which 1 is a transient, are found to be variable by using a comparison of between each exposure and its previous exposure. Although the method is slightly different than ours, we deem that the major difference is due to the higher number of observations and higher statistics. The different number of observations and sampling pattern do not allow a direct comparison. The sampling pattern is important also in the computation of the excess variance: in this work we derive $F_{\rm var}$ from the same observations, implying the same number of points and sampling pattern for all sources and allowing therefore the comparison among the variable ULXs in the Cartwheel group.

The fact that the long-term variability is found in many sources in the Cartwheel group suggests that variability on long time-scales (months to years) is a typical feature of ULX population in ring galaxies or in general disturbed by a gravitational encounter. This result agrees with the observations of ULXs in non-ring galaxies where the long-term variability is often found (e.g. \citealt{Song2020,Pintore2021}),  
while the short-term variability is suppressed in some ULXs and has a transient behaviour in others (e.g. \citealt{Gladstone2009a,Heil2009}), rendering it difficult to detect with few observations.

Variability factors of one order of magnitude or more are usually typical of the PULXs (e.g. \citealt{Israel2017,Brightman2019}). The majority of the  transient sources (5/6), corresponding to about half of the variable sample,  
have a luminosity ratio $\geq$ 10, 
suggesting that they might be PULX candidates. The large variability factors observed in the PULX are usually associated to a bi-modal flux distribution, caused by the propeller mechanism, an inhibition of accretion due to the magnetic field of a highly magnetized NS (see e.g. \citealt{Tsygankov2016a,Grebenev2017}). Thus, large variability factors and/or a transient behaviour in the long-term light-curves of the ULXs have been proposed to identify candidate PULXs (\citealt{Earnshaw2018,Song2020}). In the current study, given the small number and sparse observations, we cannot confirm when a large variability factor (or a transient behaviour) corresponds to a bi-modal flux distribution. In principle, the variation may come also from a flare in the emission, as sometimes observed in the ULXs (e.g. \citealt{Pintore2021}). Another possibility to explain large flux variations is a poorly sampled super-orbital periodicity. For example, the super-orbital periodicity in M82 X-2 was confused with a flux bi-modality (see \citealt{Tsygankov2016a,Brightman2019}).

The cumulative XLF of the Cartwheel galaxy maintains a constant shape among the three \chandra\ epochs, in contrast to the variability observed in the intensity of the individual ULXs. This suggests that the stochastic variability observed in the sources flux does not influence the average behaviour of the ULX population, which on average keeps the same luminosity distribution, linked to the star formation and the ring expansion, on the observations timescale, i.e. years, which is too short to observe an evolution in the overall population. This result indicates that the XLF may be used to characterise a population of X-ray binaries, even if they have been observed in a single epoch. Besides, we exploit the stability of the XLF to sum the different single epoch detection as different "realization" of the same measure to increase the statistical significance in order to fit the differential distribution.

The Cartwheel contains the largest number of luminous ULXs ever found in a single galaxy, as expected given the low metallicity of the environment (\citealt{Fosbury1977,ZaragozaCardiel2022}), due either to larger compact objects formed (see e.g. the discussion in \citealt{Mapelli2009,Mapelli2010}) or to an enhancement of the binary luminosity (e.g. \citealt{Fragos2013}). This is also confirmed in the recent study of \citet{Lehmer2021}.

We derive the number of ULXs per SFR: we have observed  
29 ULXs in the Cartwheel galaxy
which results in $\sim$ 1.6 ULXs/($M_{\odot}$ yr$^{-1}$) for a SFR of $\sim$ 18 $M_{\odot}$ yr$^{-1}$ (\citealt{mayya2005}). Of these 29 ULXs, 8 have shown, at least in one observation, a luminosity $>$ 10$^{40}$ erg s$^{-1}$, resulting in $\sim$ 0.4 ULXs/SFR at this luminosity level. 
The number of ULXs per SFR, with $L_{\rm x} > $10$^{39}$ erg s$^{-1}$, is consistent with the results of \citet{Lehmer2021}, for metallicity $\leq$ 8.2. 
Our result for the most luminous sources, above 10$^{40}$ erg s$^{-1}$ is consistent with the result of \citet{Lehmer2021} for metallicity $\leq$ 8.0. 

In G1 there are 7 ULXs and, assuming a SFR of $\sim$ 5 $M_{\odot}$ yr$^{-1}$ \citep{Crivellari2009}, we obtain 1.4 ULXs/($M_{\odot}$ yr$^{-1}$), a similar result to that for the Cartwheel. If instead the SFR is 10 $M_{\odot}$ yr$^{-1}$, as suggested by the XLF (see section \ref{sec:lum_func}), the number of ULXs/($M_{\odot}$ yr$^{-1}$) would reduce to 0.7, possibly indicating a metallicity $\geq$ 8.6 in G1 (see table 3 in \citealt{Lehmer2021}).

\citet{Kovlakas2020} compute the expected number of ULXs per $M_{\odot}$ yr$^{-1}$ as a function of the galaxy morphological type. The number of ULXs, i.e. $L_{\rm x}$ above 10$^{39}$ erg s$^{-1}$, (1.6 ULXs/($M_{\odot}$ yr$^{-1}$)) in the Cartwheel is smaller albeit consistent with the number of ULXs expected in late spirals/irregular galaxies from \citet{Kovlakas2020}, i.e. 2.39$^{+1.21}_{-0.91}$ ULXs/($M_{\odot}$ yr$^{-1}$).

The large number of ULXs in the Cartwheel galaxy might be due to the fact that we are looking at very young sources, considering that the star formation episode occurred a few hundreds of Myr ago (e.g. \citealt{Renaud2018,Higdon1996}), as also pointed out by \citet{wolter2018} for their sample of ring galaxies. On the other hand, the consistency of the number of ULXs we have found with the expected number of ULXs at the Cartwheel metallicity, according to the results of \citet{Lehmer2021}, suggests that a primary role is given by the the environment metallicity. Detailed analysis aimed to disentangle age and metallicity effects are on going, similar to what has been performed in the case of the collisional ring galaxy NGC 922 (\citealt{Kouroumpatzakis2021}).

The ULXs P25, P26, P30, P34 and P39 are both transient objects and have a significant variability with a luminosity ratio of $\sim$ 1 order of magnitude, rendering them the best PULXs candidates in our sample. These sources constitutes $\sim$ 15 per cent (5/37)  of the detected ULXs, suggesting that the population of ULXs in the Cartwheel group is dominated by BHs.
This hypothesis is also in line with the expectations for low metallicity environments (e.g. \citealt{Wiktorowicz2019}).

\section{Conclusion}
\label{sec:conclusion}
In this paper we study the X-ray emission of the Cartwheel galaxy and its companions, using all the available \chandra\ and \xmm\ observations, taken between 2001 and 2008. 
\begin{itemize}
\item We detect 44 sources, 29 of which positionally associated to the Cartwheel galaxy (plus two with non-significant flux detections), 8 to G1 (of which 1 possibly linked to the nucleus), 1 to G2 (possibly linked to the nucleus) and 2 to G3 (of which 1 possibly linked to the nucleus). Two sources in the area considered are not positionally associated to the galaxies in the Carthweel group and therefore we consider them interlopers.
\item No short-term variability (minutes--hours) is detected in this sample of ULXs, possibly due also to the limited statistics.
\item The long-term variability (months--years) is significant ($\geq$3$\sigma$) for 11  
sources out of 37. 
Six of these sources show a transient behaviour, five of which with 
variability factors larger than one order of magnitude, which are typical of PULX. These sources constitutes $\sim$ 15 per cent of the Cartwheel group ULXs, suggesting that this ULXs population is dominated by BHs.
\item The Cartwheel XLF is consistent in shape over the \chandra\ epochs, despite the variability of single sources, suggesting that the average behaviour of the population remains constant on years timescales. The slope of the differential XLF of the sum of the observations  
 is consistent within the uncertainties with the \citet{Grimm2003} function for HMXBs.  
 The large number of high luminosity X-ray binaries might be linked to either the age of the star forming population or the metallicity of the environment.
\item The \xmm\ observations, albeit of similar exposure times as of the \chandra\ ones, do not add substantial information to the variability, since most sources are confused.
\end{itemize} 

To improve upon these findings the only current possibility is to observe again the Cartwheel group with \chandra\ for either a deep pointing or a few medium deep ones. This would allow us to recover a larger number of sources with better statistical significance and measure variability on decades time-scales. Last but not least, a new observation could shed some light on the fate of N10, which had been one of the brightest ULX ever detected.

\section*{Acknowledgements}
This research made use of data of the \chandra\ (NASA) and \xmm\ (ESA) observatories.

In this work we used the software {\sc ciao} \citep{Fruscione2006} and {\sc ds9} \citep{Joye2011} by the \textit{Chandra X-ray Center} (CXC), {\sc xspec} \citep{Arnaud1996}, {\sc sas} \citep{SAS2014}, {\sc scipy} \citep{SciPy-NMeth2020}. 

This work has been partially supported by the ASI - INAF agreement 2017-14-H.0.

We thank the anonymous referee for the many detailed comments that greatly helped improving the manuscript.

\section*{Data availability}
The data used in this work are publicly available from the NASA HEASARC archive: \url{ https://heasarc.gsfc.nasa.gov/docs/archive.html}. 
The {\sc blike} library and the scripts used for the variability analysis are available at \url{https://github.com/andrea-belfiore/BLike} (A. Belfiore, in prep.).




\bibliographystyle{mnras}
\bibliography{prova_biblio}{} 



\appendix

\section{Point sources information}

\label{appendixA}

\begin{table*}
\begin{center} 
\begin{tabular}{ccccccccccc} \hline
{Source} & {\em cts$_{2019}$} & {\em L$_{x,2019}$}& {\em cts$_{9531}$} & {\em L$_{x,9531}$} &  {\em cts$_{9807}$} & {L$_{x,9807}$} & {\em RA} & {\em Dec} & {\em Det} & {\em Sig}\\
(1) & (2) & (3) & (4) & (5) & (6) & (7) & (8) & (9) & (10) & (11)\\
\hline
N1$^{*}$ &  & &  & &  & & 0:37:45.3 & $-$33:42:28.7 & 1,2,3,m & 31.0\\
N2(G1) & 34.6$\pm$7.0 & {\it 4.8$\pm$0.8} & 21.2$\pm$5.8 & {\it 4.8$\pm$1.1} & 14.4$\pm$5.0 & {\it 3.3$\pm$0.9} & 0:37:43.9 & $-$33:42:09.9 & 1,2,3,m & 21.7\\
N3(G1) & 53.7$\pm$8.5 & {\it 7.0$\pm$0.9} & 68.9$\pm$9.4 & {\it 16.0$\pm$1.9} & 50.7$\pm$8.3 & {\it 11.5$\pm$1.6} & 0:37:43.1 & $-$33:42:04.1 & 1,2,3,m & 58.4\\
N4(G1) & 42.7$\pm$7.7 & {\it 5.7$\pm$0.9} & 15.9$\pm$5.2 & {\it 3.7$\pm$0.9} & 18.7$\pm$5.6 & {\it 4.5$\pm$1.1} & 0:37:43.0 & $-$33:42:06.1 & 1,3,m & 27.5\\
N5(G1) & 35.4$\pm$7.1 & {\it 4.6$\pm$0.8} & 5.7$\pm$3.8 & & 8.2$\pm$4.1 & {\it 1.8$\pm$0.7} & 0:37:42.8 & $-$33:42:12.7 & 1,3,m & 17.1\\
N6 & 16.1$\pm$5.2 & \underline{2.2$\pm$0.5} & 2.0$\pm$2.9 & & 3.3$\pm$3.2 & & 0:37:42.5 & $-$33:43:04.1 & 1,m & 8.8\\
N7 & 57.0$\pm$8.7 & \underline{7.8$\pm$1.0} & 9.0$\pm$4.3 & \underline{2.0$\pm$0.7} & 5.6$\pm$3.8 & \underline{1.5$\pm$0.7} & 0:37:41.1 & $-$33:43:31.8 & 1,m & 25.7\\
N8$^{*}$ & & &  & &  & & 0:37:41.1 & $-$33:42:21.5 & 1,3,m & 12.4\\
N9 & 56.8$\pm$8.7 & \underline{7.2$\pm$0.9} & 50.6$\pm$8.3 & \underline{11.3$\pm$1.6} & 23.7$\pm$6.1 & \underline{5.8$\pm$1.2} & 0:37:40.9 & $-$33:43:31.0 & 1,2,3,m & 43.8\\
N10 & 342.9$\pm$19.6 & \underline{42.8$\pm$2.3} & 150.1$\pm$13.3 & \underline{33.6$\pm$2.7} & 45.8$\pm$7.9 & \underline{11.0$\pm$1.6} & 0:37:39.4 & $-$33:43:23.3 & 1,2,3,m & 132.6\\
N11 & 60.1$\pm$8.9 & \underline{7.8$\pm$1.0} & 33.2$\pm$6.9 & \underline{7.6$\pm$1.3} & 26.2$\pm$6.3 & \underline{6.0$\pm$1.2} & 0:37:39.2 & $-$33:42:50.3 & 1,2,3,m & 38.4\\
N12 & 39.1$\pm$7.4 & \underline{5.0$\pm$0.8} & 21.3$\pm$5.8 & \underline{4.8$\pm$1.1} & 9.2$\pm$4.3 & \underline{2.1$\pm$0.7} & 0:37:39.2 & $-$33:42:29.7 & 1,2,3,m & 23.4\\
N13 & 32.2$\pm$6.8 & \underline{4.4$\pm$0.7} & 21.5$\pm$5.8 & \underline{5.5$\pm$1.1} & 22.7$\pm$6.0 & \underline{6.1$\pm$1.2} & 0:37:38.8 & $-$33:43:18.8 & 1,2,3,m & 24.0\\
N14 & 77.2$\pm$9.9 & \underline{9.6$\pm$1.1} & 39.5$\pm$7.4 & \underline{9.1$\pm$1.4} & 46.7$\pm$8.0 & \underline{11.5$\pm$1.7} & 0:37:38.7 & $-$33:43:16.3 & 1,2,3,m & 53.4\\
N15 & 15.5$\pm$5.1 & \underline{2.0$\pm$0.5} & 9.8$\pm$4.4 & \underline{2.2$\pm$0.7} & 5.6$\pm$3.6 & 1.4$\pm$0.6 & 0:37:38.4 & $-$33:43:08.9 & 1,2,m & 10.4\\
N16 & 83.4$\pm$10.3 & \underline{11.7$\pm$1.2} & 33.5$\pm$6.9 & \underline{7.7$\pm$1.3} & 47.5$\pm$8.0 & \underline{11.3$\pm$1.6} & 0:37:37.6 & $-$33:42:55.2 & 1,2,3,m & 51.8\\
N17 & 92.8$\pm$10.7 & \underline{12.8$\pm$1.3} & 37.4$\pm$7.2 & \underline{8.8$\pm$1.4} & 16.3$\pm$5.2 & \underline{3.9$\pm$1.0} & 0:37:37.6 & $-$33:42:57.0 & 1,2,m & 32.7\\
N18 & 6.3$\pm$3.8 & 0.8$\pm$0.3 & 1.9$\pm$2.9 & & 1.5$\pm$2.7 & & 0:37:43.4 & $-$33:43:12.8 & m & 3.9\\
N19(G1) & 10.9$\pm$4.6 & {\it 1.6$\pm$0.5} & 14.4$\pm$5.0 & {\it 3.3$\pm$0.9} & 7.3$\pm$4.0 & {\it 1.7$\pm$0.6} & 0:37:42.8 & $-$33:42:09.8 & 1,2,m & 12.1\\
N20$^{*}$ & 5.5$\pm$3.6 & 1.1$\pm$0.4 & 0 & & 0 & & 0:37:42.1 & $-$33:43:13.8 & 1 & \\
N21 & 6.0$\pm$3.8 & 0.8$\pm$0.4 & 15.3$\pm$5.1 & \underline{3.7$\pm$0.9} & 6.0$\pm$3.8 & \underline{1.4$\pm$0.6} & 0:37:41.2 & $-$33:42:32.4 & 2,m & 10.1\\
N22 & 13.4$\pm$5.0 & \underline{1.7$\pm$0.5} & 8.2$\pm$4.3 & \underline{1.9$\pm$0.7} & 6.4$\pm$4.0 & \underline{1.6$\pm$0.7} & 0:37:40.5 & $-$33:43:24.7 & 1,3,m & 10.1\\
N23 & 10.3$\pm$4.6 & 1.3$\pm$0.4 & 0 & & 0.7$\pm$2.3 & & 0:37:42.0 & $-$33:43:26.9 & 1,m & 3.4\\
N24$^{*}$ & 5.3$\pm$3.8 & & 2.5$\pm$3.2 & & 3.3$\pm$3.4 & &  0:37:40.2 & $-$33:43:26.9 & 1 &\\
P25 & 1.9$\pm$2.9 & & 29.1$\pm$6.6 & \underline{6.5$\pm$1.2} & 21.3$\pm$5.8 & \underline{5.1$\pm$1.1} & 0:37:42.4 & $-$33:42:49.8 & 2,3,m & 20.5\\
P26$^{\dagger}$ & 1.9$\pm$3.2 & & 2.1$\pm$2.9 & & 29.4$\pm$6.6 & \underline{7.3$\pm$1.4} & 0:37:40.6 & $-$33:43:28.4 & 3,m & 12.0\\
P27$^{\dagger}$ & 22.9$\pm$6.0 & \underline{3.0$\pm$0.6} & 12.1$\pm$4.7 & \underline{3.0$\pm$0.8} & 3.8$\pm$3.4 & & 0:37:39.4 & $-$33:43:21.2 & m & 11.0\\
P28$^{\dagger}$ & 1.8$\pm$2.9 & & 7.0$\pm$4.0 & \underline{1.8$\pm$0.7} & 6.4$\pm$3.8 & \underline{1.6$\pm$0.6} &  0:37:38.1 & $-$33:43:05.6 & m & 6.3\\
P29$^{\dagger}$ & 4.3$\pm$3.4 & & 2.1$\pm$2.9 & & 8.3$\pm$4.1 & \underline{1.9$\pm$0.7} & 0:37:42.9 & $-$33:43:14.4 & 3,m & 5.8\\
P30$^{\dagger}$(G1) & 1.2$\pm$2.7 &  & 15.0$\pm$5.1 & {\it 3.4$\pm$0.9} & 2.4$\pm$2.9 & & 0:37:43.8 & $-$33:42:14.4 & 2,m & 6.6\\
P31(G1) & 29.7$\pm$6.6 & {\it 3.8$\pm$0.7} & 14.9$\pm$5.1 & {\it 3.5$\pm$0.9} & 15.7$\pm$5.2 & {\it 3.6$\pm$0.9} &  0:37:42.9 & $-$33:42:04.2 & 1,m & 22.1\\
P32(G3) & 8.3$\pm$4.1 & 1.1$\pm$0.4 & 7.3$\pm$4.0 & 1.8$\pm$0.7 & 5.8$\pm$3.6 & 1.9$\pm$0.7 & 0:37:47.0 &  $-$33:39:53.3 & 1,2,3,m & 12.1\\
P33$^{\dagger}$ & 1.0$\pm$2.7 & & 6.2$\pm$3.8 & \underline{1.4$\pm$0.6} & 1.8$\pm$2.9 & & 0:37:40.7 & $-$33:42:58.3 & m & 3.8\\
P34$^{\dagger}$ & 0 & & 0 & & 9.1$\pm$4.3 & \underline{2.2$\pm$0.7} & 0:37:38.7 & $-$33:43:13.2 & 3,m & 3.5\\
P35$^{\dagger}$ & 3.9$\pm$3.6 & & 6.7$\pm$4.0 & \underline{1.5$\pm$0.6} & 3.5$\pm$3.4 & & 0:37:41.3 & $-$33:43:31.5 & m & 6.7\\
P36 & 8.8$\pm$4.3 & 1.2$\pm$0.4 & 4.8$\pm$3.6 & & 0.6$\pm$2.3 & & 0:37:37.9 & $-$33:42:53.1 & m & 4.7\\
P37 & 9.2$\pm$4.3 & 1.2$\pm$0.4  & 2.4$\pm$2.9 & & 0.2$\pm$2.3 & & 0:37:41.7 & $-$33:42:35.3 & 1,m & 5.9\\
P38$^{\dagger}$ & 10.3$\pm$4.5 & 1.3$\pm$0.4 & 2.5$\pm$3.2 & & 9.3$\pm$4.4 & \underline{2.5$\pm$0.8} & 0:37:40.3 & $-$33:43:27.1 & 3,m & 10.0\\
P39$^{\dagger}$ & 0 & & 6.3$\pm$3.8 & \underline{1.5$\pm$0.6} & 0.2$\pm$2.3 &  & 0:37:41.0 & $-$33:42:39.6 & 2,m & 2.8\\
P40$^{\dagger}$ & 0.7$\pm$2.7 & & 0.4$\pm$2.3 & & 7.1$\pm$4.0 & \underline{1.6$\pm$0.6} & 0:37:41.8 & $-$33:42:43.3 & 3,m & 3.0\\
P41$^{\dagger}$ & 0.1$\pm$2.3 & & 4.1$\pm$3.4 & & 2.6$\pm$2.9 & & 0:37:38.5 & $-$33:42:57.1 & m & 2.9\\
P42$^{\dagger}$ (G1n) & 5.0$\pm$3.6 &  & 0.3$\pm$2.3 &  & 2.2$\pm$2.9 &  & 0:37:43.5 & $-$33:42:09.0 & m & 4.1\\
P43$^{\dagger}$ (G2n) & 6.9$\pm$4.0 & 0.9$\pm$0.4 & 10.4$\pm$4.4 & 2.6$\pm$0.8 & 6.8$\pm$4 & 1.6$\pm$0.6 & 0:37:44.9 & $-$33:42:20.3 & 2,m & 10.0\\
P44$^{\dagger}$ (G3n) & 10.9$\pm$4.8 & 1.4$\pm$0.5 & 3.5$\pm$3.6 & & 4.4$\pm$3.6 &  & 0:37:46.3 & $-$33:39:54.0 & 1,m & 5.0\\
\hline
\end{tabular}
\end{center}
\caption{Column 1: source name; with G1n, G2n and G3n we indicate three sources probably associated with the nuclear regions of the companion galaxies. 
We mark with simbol $^{\dagger}$ the new sources from this work, not published in previous analysis of the Cartwheel group. Columns 2,4,6: net counts (from {\sc xspec}) of the sources found in the Cartwheel Galaxy and in the companion galaxies G1, G2, G3, in the energy band 0.3-10 keV. The errors are computed using the \citet{Gehrels1986} approximation. Columns 3, 5, 7: luminosity in 2-10 keV, in units of 10$^{39}$ erg s$^{-1}$, for the sources in the Cartwheel group with a significant flux determination ($>$ 3$\sigma$ with {\sc blike}).  
We underline the luminosities used in the computation of the Cartwheel differential XLF: the luminosities $\geq$ 1.4$\times$10$^{39}$ erg s$^{-1}$ of the sources detected in the Cartwheel. We list in italic those used for the G1 cumulative XLF.  The 2-10 keV luminosities have been derived from table \ref{tab:lmax_lmin}, by dividing the 0.5-10 keV luminosities by a conversion factor of 1.6, corresponding to the (0.5-10)keV/(2-10)keV luminosities from the average spectrum. 
Column 8,9: coordinates for each source obtained by running {\sc wavdetect} in the merged \chandra\ image, with a significance $>$ 2$\sigma$, in the energy band 0.3-10 keV. 
$^{(*)}$Source N1 is outside the galaxy area, even if close to G2. Source N8 is in the region between the Cartwheel and G1 and not spatially consistent with any of the group galaxies. Therefore we just report the coordinates and detection details for these sources, and we do not analyse them further.  
$^{(*)}$ N20 and N24 are not detected in the merged image, but just in observation 2019, we thus report the coordinates from observation 2019. Column 10: observations in which a source is detected with {\sc wavdetect}; 1: 2019, 2: 9531, 3: 9807, m: merge. Column 11: detection significance in the merged image, in $\sigma$ units. 
}
\label{tab:netcont}
\end{table*}

\begin{table}
\begin{center} 
\begin{tabular}{ccc} \hline
{\em } & {\em 9531-2019} & {\em 9807-2019} \\ 
\hline
$\alpha$ & $-$0.158 & $-$0.099\\
$\delta$ &  0.100 & $-$0.078\\
\hline
\end{tabular}
\end{center}
\caption{Average of the differences between right ascension ($\alpha$) and declination ($\delta$) of the observations 9531 and 9807, with respect to observation 2019. The differences are expressed in arcseconds.}
\label{tab:diff_coord}
\end{table}

The \chandra\ coordinates of the detected sources derived from the merged image, as well as their net counts in the three \chandra\ exposures are reported in table \ref{tab:netcont}. The data products are extracted from regions with 1 arcsec radius, as the best compromise to both have a large Enclosed Energy Fraction ($\sim$ 90 per cent) and to include just a single detected source in the extraction region, considering the closeness of some sources both in the Cartwheel ring and in the galaxy G1. For each source we use a different background, an elliptical annulus around the source,
because the diffuse background emission is not uniform across the
galaxy \citep{Crivellari2009}. In particular, the Cartwheel ring has
more diffuse gas than the other regions of the galaxy. We use the same nomenclature as in \citet{Wolter2004} for sources that we have matched in common with that study and "P", with progressive numbers, for "new" sources, not detected in  \citet{Wolter2004} (with the exception of P32 which is present in \citealt{Wolter2004} and was labeled G3 in Table 1 of Appendix A). 
The matching between the coordinates listed in \citet{Wolter2004} and those detected in the present work, has been done by visually inspecting the \chandra\ images. We also derive the difference in our coordinates with respect to those of \citet{Wolter2004} and consider a match true when it is smaller than the PSF. 

Given that the astrometry is in principle different for the three
pointings, we compare the source positions in the three observations.
We compute the differences in right ascension and declination with
respect to obsId 2019: the resulting average values (in arcseconds)
are shown in Table \ref{tab:diff_coord}. The values are smaller than the minimum PSF
in the region of interest (which is $\sim$ 1 arcsec in radius) so we can
neglect these differences in our analysis. 

In order to derive the number of spurious sources in the area occupied by the Cartwheel galaxy, we consider an elliptical region, centered on coordinates RA=0:37:40.5, Dec=$-$33:42:59.0 and with semi-minor and semi-major axes of 35 and 45 arcsec, which includes the whole galaxy region. The area of the Cartwheel galaxy is $\sim$ 0.0004 deg$^{2}$. We derive the number of expected extragalactic sources by square degree from the LogN-LogS relation of \citet{Moretti2003}. In the Cartwheel area we expect $\sim$ 0.4 contaminants in the energy band (2-10) keV. Considering all the galaxies in the Cartwheel group, the expected number of interlopers remains smaller than 1 ($\sim$ 0.6).

\section{Likelihood function}
\label{appendixB}
We report in this section the derivation of the likelihood function for the power-law model. An analogous procedure has been used to derive the likelihood function for the cut-off power-law model.
\newline
\newline
We construct a binned differential luminosity function, grouping the sources in equally logarithmic spaced bins, with a step log($\Delta$) = 0.3, the step has been chosen not to have empty bins, obtaining five bins. It follows that $x_{i}$=$x_{0}\times \Delta^{i}$, where $x_{0}$ and $x_{i}$ are the lower luminosity bounds of the first and i-th bin, respectively, in units of erg s$^{-1}$. 
We assume that the XLF has a power-law shape $f$=$n\, x^{-\gamma}$, with $n$  the normalization, $\gamma$ the power-law index and $x$ the luminosity in units of 10$^{39}$ erg s$^{-1}$.

Every bin of the XLF contains $N_{i}$ sources and the expected value for each bin is:

\begin{equation}
   y_{i}=\int_{x_{i}}^{x_{i+1}}{nx^{-\gamma}dx} = \frac{n}{1-\gamma} (x_{i+1}^{1-\gamma} - x_{i}^{1-\gamma}) = \frac{n \times (\Delta^{1-\gamma} - 1)}{1-\gamma} x_{i}^{1-\gamma}
\label{eq:yi}
\end{equation}
and dividing by the bin width ($l_{i}$ = $x_{i+1}$ - $x_{i}$) we obtain the desired value:
\begin{equation}
    \frac{y_{i}}{l_{i}} = \frac{n(\Delta^{1-\gamma}-1)x_{i}^{-\gamma}}{(1-\gamma)(\Delta-1)}.
    \label{eq:yidL}
\end{equation}
Introducing
\begin{equation}
    \alpha(n,\gamma) = \frac{n(\Delta^{1-\gamma}-1)}{(1-\gamma)(\Delta-1)}
    \label{eq:alpha}
\end{equation}
we obtain the more compact form for the expected value:
\begin{equation}
    \frac{y_{i}}{l_{i}} = \alpha x_{i}^{-\gamma}.
\label{eq:yidLalpha}
\end{equation}
The sources distribution in the bins follows the Poisson statistics, so the probability to have $N_{i}$ sources in each bin, given the expected value $y_{i}$ is:
\begin{equation}
    P(N_{i}|y_{i}) = e^{-y_{i}} \frac{y_{i}^{N_{i}}}{N_{i}!}    
\end{equation}
which is the likelihood function in the single bin. The factorial term can be neglected because it is model independent.

The total probability is the product of the probabilities of the single bins. To simplify the calculation we consider the sum of the logarithms of the probability. 
\begin{equation}
\ln(P_{\rm tot}) = \ln\left(\prod  P(N_{i}|y_{i})\right) = \displaystyle\sum_{i} \ln(e^{-y_{i}} y_{i}^{N_{i}}) = \displaystyle\sum_{i} (-y_{i} + N_{i} \ln(y_{i}))
\label{eq:lnP}
\end{equation}
and, using $\frac{N_{i}}{l_{i}}$ and $\frac{y_{i}}{l_{i}}$ instead of $N_{i}$ and $y_{i}$, from equations \ref{eq:lnP} and \ref{eq:yidLalpha} we write the log likelihood:
\begin{equation}
    \ln(Like) = -\alpha \displaystyle\sum_{i} x_{i}^{-\gamma} + \ln \alpha \displaystyle\sum_{i} \frac{N_{i}}{l_{i}} - \gamma \displaystyle\sum_{i} \frac{N_{i}}{l_{i}} \ln x_{i}.
\end{equation}

With the same procedure, we calculate the likelihood for an exponential cut-off power-law $y$ = $nx^{-\gamma} e^{-\frac{x}{x_c}}$, where $x_c$ is the cut-off luminosity in units of 10$^{39}$ erg s$^{-1}$. In this case, the logarithmic likelihood function has the form:
\begin{equation}
    \ln(Like) = \displaystyle\sum_{i} \left [\frac{n}{x_{i+1}-x_{i}}*p + \frac{N_{i}}{l_{i}} \ln \left(\frac{-n}{x_{i+1} - x_{i}}*p \right)\right ]
\end{equation}
with 
\begin{equation}
    p = \left(x_{i+1}^{1-\gamma}\right) E_{\gamma}\left(\frac{x_{i+1}}{x_{c}}\right) - \left(x_{i}^{1-\gamma}\right) E_{\gamma}\left(\frac{x_{i}}{x_{c}}\right)
\end{equation}
where $E_{\gamma}$ is the exponential integral.

\section{Diffuse emission}
\label{appendixC}

\begin{table*}
\begin{center} 
\begin{tabular}{cccccccc} \hline
{\rm Src} & {$n_{\rm H}$} & {\it $\it \Gamma$} & {\em $kT$} & {$\chi^{2}$/dof} & {$L_{\rm gas}$} & {$L_{\rm pow}$} & {$L_{\rm tot}$} \\ 
        &   { 10$^{21}$ cm$^{-2}$} & & { keV} & & \multicolumn{3}{c}{$10^{40}$ erg s$^{-1}$} \\
\hline
Cartwheel &  1.2$^{+1.8}_{-1.1}$ & 2.3$^{+1.3}_{-0.8}$ & 0.25$^{+0.11}_{-0.06}$ & 42.2/57 & 1.3$\pm$2.6 & 3.3$\pm$1.9 & 24.7$\pm$3.3\\

G1 &  1.6$^{+2.7}_{-1.6}$ & 2.3$^{+1.1}_{-0.7}$ & -- & 15.4/21 & -- & 1.4$\pm$0.2 &  6.9$\pm$0.5\\
G2 &   1.2 & 1.5$^{+1.4}_{-1.5}$ & -- & 14.2/14 & -- & 0.7$\pm$0.4 & 1.1$\pm$0.4\\ 
G3 &  1.2 & 3.4$^{+1.3}_{-1.1}$ & -- & 20.6/29 & -- & 0.5$\pm$0.2 & 0.9$\pm$0.2\\
\hline
\end{tabular}
\end{center}
\caption{Spectral parameters from the fit of the non-resolved component for the Cartwheel galaxy, G1, G2 and G3. See text for more details. The model used is {\sc tbabs*(apec+pow)} in {\sc xspec} for the Cartwheel and {\sc tbabs*pow} for the companion galaxies. The unabsorbed luminosities are in units of 10$^{40}$ erg s$^{-1}$. The gas luminosity (from the {\sc apec} component) is given in (0.5-2) keV, while the power-law luminosities are in (0.5-10) keV. The luminosities in the table are average values among the three \chandra\ epochs. In the last column we report the total luminosity for each galaxy, derived as the sum of the non-resolved emission plus the sum of the point-like resolved emission (from the luminosities in table \ref{tab:lmax_lmin}). For the resolved emission we use the maximum luminosity observed, i.e. C.: obs. 2019, G1: obs. 9531, G2: obs. 9531, G3: obs. 2019.} 
\label{tab:fit_g1_g2_g3}
\end{table*}

The Cartwheel Galaxy shows also diffuse emission, due to unresolved point-like sources and diffuse gas. 
We extract a spectrum from the three \chandra\ images using an elliptical region centered on the coordinates RA = 0:37:40.5 and Dec = --33:42:59.0, with major radius = 45 arcsec 
and minor radius = 35 arcsec, including both the ring and the region inside the galaxy, but excluding the detected point-like sources. 

\begin{table}
\begin{center} 
\begin{tabular}{cccccc} \hline
{Source} & {DEC} & {RA} & {major axis} & {minor axis} & {ref.} \\ 
\hline
G1 & 0:37:43.5 & -33:42:06.5 & 15 & 11 & 1\\
G2 & 0:37:45.0 & -33:42:20.5 & 12 & 12 & 2\\  
G3 & 0:37:46.1 & -33:39:52.1 & 26 & 14 & 2\\
\hline
\end{tabular}
\end{center}
\caption{This table shows the coordinates of the galaxies G1, G2 and G3. We extract their spectra using elliptical regions, with axis indicated in the fourth and fifth columns of the table in arcseconds. The coordinates are taken from: (1) \citet{Skrutskie2006}; (2) \citet{Gaia2020}.}
\label{tab:g1_g2_g3}
\end{table}
We group the data in order to have at least 2$\sigma$ significance in each bin. As also found from previous analysis (\citealt{Wolter2004}), a single power-law gives an acceptable fit but the residuals improve with an additional component. We thus use a plasma model plus a power-law to fit the data, multiplied by an absorption component ({\sc tbabs*(apec+pow)} in {\sc xspec}). We fix the abundances in the plasma model at 0.5$\times$solar value, considering the low metallicity of the Cartwheel galaxy (\citealt{Fosbury1977,ZaragozaCardiel2022}). The plasma component models the diffuse gas, while the power-law may represent the unresolved point-like sources. We first fit the three \chandra\ observations, linking the $n_{\rm H}$ and leaving the other parameters free to vary independently among the observations. Even if there are differences among the best-fitting parameters, suggesting a possible variability in the spectral shape, the uncertainties are large and the parameters are consistent within the errors among the observations. Therefore, we decide to link also the other parameters to better constrain them, but we add a multiplicative constant to our model to take into account possible differences in flux among the observations. The best-fitting parameters are shown in Table \ref{tab:fit_g1_g2_g3}.
The unabsorbed fluxes obtained are consistent among the observations within the errors. The average values found are: $F_{\rm gas}^{(0.5-2)\rm keV}$ = 6.9$\times$10$^{-15}$ erg s$^{-1}$ cm$^{-2}$; $F_{\rm pow}^{(0.5-2)\rm keV}$ = 9.8$\times$10$^{-15}$ erg s$^{-1}$ cm$^{-2}$;  $F_{\rm pow}^{(2-10)\rm keV}$ = 7.4$\times$10$^{-15}$ erg s$^{-1}$ cm$^{-2}$. The gas component peaks below 2 keV and therefore contributes only to the soft band, for a yield similar to the unresolved sources, which however have a comparable flux also in the harder band.
While the power-law index and the gas temperature found are similar to those obtained by \citet{Wolter2004}, our best-fitting $n_{\rm H}$ is smaller, which causes the unabsorbed flux derived for the first \chandra\ observation, especially for the gas component, to be smaller than the value obtained in the previous analysis. 
However, given the large uncertainties our results are broadly consistent with those already published. 
The gas contributes to $\sim$ 30 per cent of the diffuse emission in the total energy band (the contribution of the gas is just below 2 keV, i.e. at soft energies) and to $\sim$ 40 per cent in the soft band, 
while the corresponding contributions due to unresolved sources are $\sim$ 70 and 60 per cent, in the total and soft energy bands respectively. The non-resolved emission contributes $\sim$ 20 per cent of the total emission of the Cartwheel.

As already found in previous works (\citealt{Wolter2004,Crivellari2009}) and confirmed with our spectral analysis, which considers all the three \chandra\ epochs, the diffuse emission is well modelled with two components, compatible with emission from an unresolved population of X-ray binaries and from a diffuse and hot gas. Considering the large number of ULXs on the ring, it is reasonable to expect a population of less luminous binaries, generated in the same episode of star formation, but too faint to be resolved with the currently active X-ray satellites at the Cartwheel distance. 
Considering the Cartwheel observations and the results from the simulations \citep{Renaud2018}, the Cartwheel galaxy appears in a phase with the star formation (SF) concentrated in the outer ring. The simulations predict that the shock-wave, generated after the impact between the Cartwheel and the intruder galaxy, activate at first the SF just in the outer ring. The gas in the outer ring is then expected to flow towards the internal regions of the galaxy, through the spokes. As a result, the activation of the SF moves in time from the outer ring to the inner regions of the galaxy. This would imply an age from the impact smaller than 100 Myr, according to the simulations. From other considerations, based on dynamical arguments, a larger age ($>$ 200 Myr) is derived (\citealt{Higdon1996,Amram1998}). If the SF has been activated just in the outer ring, but not in the inner galaxy regions, we would expect to see both unresolved point sources and diffuse hot plasma in the outer ring, but only hot plasma emission in the inner regions. The statistics of the \chandra\ data is not enough to separate the diffuse emission of the inner regions from the one of the outer ring. \citet{Crivellari2009} used \xmm\ data to study the diffuse emission in the two different regions, thanks to their higher statistics. 
Unexpectedly, they found two components also in the inner galaxy regions. Although, as they suggested, a power-law-like contribution from the PSF wings of the brightest sources is possible, this might be also an indication of the presence of SF also in the inner ring, maybe due to a different phase of the shock wave.
To study separately the diffuse emission in the two indicated regions, an instrument with the spatial resolution of \chandra\ and the \xmm\ effective area would be ideal. At present, the goal may be reached with a longer \chandra\ exposure (of at least 100-200 ks) 
that would provide enough statistics for a more detailed analysis of the diffuse emission.
\newline
\newline
To study the non-resolved emission of the three companion galaxies, 
we use the elliptical regions indicated in Table \ref{tab:g1_g2_g3}, excluding the resolved point-like sources.

For G1, we bin data in order to have at least 20 counts in each bin and use the model {\sc tbabs*power-law}. The fit parameters values are consistent in the three observations within the errors. Therefore we analyse the three spectra together, in order to increase the statistics. The best-fitting parameters are given in Table \ref{tab:fit_g1_g2_g3}. The non-resolved emission is $\sim$ 20 per cent of the total emission of the galaxy.

We repeat the same procedure for G2 and G3. For these galaxies, we have not enough statistics to use complex models and indeed to fit more than one parameter, so we fixed the $n_{\rm H}$ at 1.2$\times$10$^{21}$ cm$^{-2}$, the value derived from the average spectrum of the point-like sources, while we leave the power-law index free.

The best-fitting power-law index and the unabsorbed luminosity for G2 and G3 are indicated in Table \ref{tab:fit_g1_g2_g3}. From the same table we can see that the non-resolved emission constitutes $\sim$ 64 and 56 per cent of the total emission, respectively for G2 and G3.


\bsp	
\label{lastpage}
\end{document}